\documentclass[longauth]{aa}
\usepackage{graphicx}
\usepackage{rotating}
\usepackage{adjustbox}
\usepackage{geometry}
\usepackage{natbib}
\usepackage{txfonts}
\usepackage{rotating}
\usepackage{pdflscape}
\usepackage{subcaption}
\usepackage{xcolor}
\usepackage{threeparttable}
\usepackage{hyperref}
\usepackage{lastpage}
\usepackage{orcidlink}
\usepackage{tabularx}
\newcolumntype{C}{>{\centering\arraybackslash}X}

\begin{document} 
\bibliographystyle{aa}

   \title{Super-Earth masses and stellar abundances from NIRPS reveal tentative evidence for water-rich formation around M dwarfs}

    \titlerunning{Results from the NIRPS CMF subprogram}
    \authorrunning{Weisserman et al.}
   
\author{
Drew Weisserman\inst{1,*}\orcidlink{0000-0002-7992-469X},
Nicole Gromek\inst{1}\orcidlink{0009-0000-1424-7694},
Ryan Cloutier\inst{1}\orcidlink{0000-0001-5383-9393},
Komal Bali\inst{2}\orcidlink{0000-0003-4829-7141},
Charles Cadieux\inst{3}\orcidlink{0000-0001-9291-5555},
Mykhaylo Plotnykov\inst{4}\orcidlink{0000-0002-9479-2744},
Alexandrine L'Heureux\inst{3}\orcidlink{0009-0005-6135-6769},
Avidaan Srivastava\inst{3}\orcidlink{0009-0009-7136-1528},
Andres Carmona\inst{5}\orcidlink{0000-0003-2471-1299},
Yolanda G. C. Frensch\inst{6,7,}\orcidlink{0000-0003-4009-0330}
% ...
}

\institute{
\inst{1}Department of Physics \& Astronomy, McMaster University, 1280 Main St W, Hamilton, ON, L8S 4L8, Canada\\
\inst{2}Institute for Particle Physics and Astrophysics, ETH Z\"urich, Otto-Stern-Weg 5, 8093 Z\"urich, Switzerland\\
\inst{3}Institut Trottier de recherche sur les exoplan\`etes, D\'epartement de Physique, Universit\'e de Montr\'eal, Montr\'eal, Qu\'ebec, Canada\\
\inst{4}Department of Physics, University of Toronto, Toronto, ON M5S 3H4, Canada\\
\inst{5}Univ. Grenoble Alpes, CNRS, IPAG, F-38000 Grenoble, France\\
\inst{6}Observatoire de Gen\`eve, D\'epartement d’Astronomie, Universit\'e de Gen\`eve, Chemin Pegasi 51, 1290 Versoix, Switzerland\\
\inst{7}European Southern Observatory (ESO), Av. Alonso de Cordova 3107,  Casilla 19001, Santiago de Chile, Chile\\
\inst{8}Observatoire du Mont-M\'egantic, Qu\'ebec, Canada\\
\inst{9}Instituto de Astrof\'isica e Ci\^encias do Espa\c{c}o, Universidade do Porto, CAUP, Rua das Estrelas, 4150-762 Porto, Portugal\\
\inst{10}Departamento de F\'isica e Astronomia, Faculdade de Ci\^encias, Universidade do Porto, Rua do Campo Alegre, 4169-007 Porto, Portugal\\
\inst{11}Department of Earth, Planetary, and Space Sciences, University of California, Los Angeles, CA 90095, USA\\
\inst{12}Department of Physics, McGill University, 3600 rue University, Montr\'eal, QC, H3A 2T8, Canada\\
\inst{13}Department of Earth \& Planetary Sciences, McGill University, 3450 rue University, Montr\'eal, QC, H3A 0E8, Canada\\
\inst{14}Centre Vie dans l’Univers, Facult\'e des sciences de l’Universit\'e de Gen\`eve, Quai Ernest-Ansermet 30, 1205 Geneva, Switzerland\\
\inst{15}Instituto de Astrof\'isica de Canarias (IAC), Calle V\'ia L\'actea s/n, 38205 La Laguna, Tenerife, Spain\\
\inst{16}Departamento de Astrof\'isica, Universidad de La Laguna (ULL), 38206 La Laguna, Tenerife, Spain\\
\inst{17}Departamento de F\'isica Te\'orica e Experimental, Universidade Federal do Rio Grande do Norte, Campus Universit\'ario, Natal, RN, 59072-970, Brazil\\
\inst{18}European Southern Observatory (ESO), Karl-Schwarzschild-Str. 2, 85748 Garching bei M\"unchen, Germany\\
\inst{19}Space Research and Planetary Sciences, Physics Institute, University of Bern, Gesellschaftsstrasse 6, 3012 Bern, Switzerland\\
\inst{20}Consejo Superior de Investigaciones Cient\'ificas (CSIC), E-28006 Madrid, Spain\\
\inst{21}Bishop's Univeristy, Dept of Physics and Astronomy, Johnson-104E, 2600 College Street, Sherbrooke, QC, Canada, J1M 1Z7, Canada\\
\inst{22}Department of Physics, Engineering Physics, and Astronomy, Queen’s University, 99 University Avenue, Kingston, ON K7L 3N6, Canada\\
\inst{23}Department of Physics and Space Science, Royal Military College of Canada, 13 General Crerar Cres., Kingston, ON K7P 2M3, Canada\\
\inst{24}Centro de Astrobiolog\'ia (CAB), CSIC-INTA, Camino Bajo del Castillo s/n, 28692, Villanueva de la Ca\~nada (Madrid), Spain\\
\inst{25}Plan\'etarium de Montr\'eal, Espace pour la Vie, 4801 av. Pierre-de Coubertin, Montr\'eal, Qu\'ebec, Canada\\
\inst{26}Division of Astrophysics, Department of Physics, Lund University, Box 118, SE-22100 Lund, Sweden\\
\inst{27}Light Bridges S.L., Observatorio del Teide, Carretera del Observatorio, s/n Guimar, 38500, Tenerife, Canarias, Spain\\
\inst{28}University Observatory, Faculty of Physics, Ludwig-Maximilians-Universit\"at M\"unchen, Scheinerstr. 1, 81679 Munich, Germany\\
\inst{29}Department of Physics, The University of Warwick, Gibbet Hill Road, Coventry, CV4 7AL, UK\\
\inst{30}Department of Astronomy \& Astrophysics, University of Chicago, 5640 South Ellis Avenue, Chicago, IL 60637, USA\\
\inst{31}Laboratoire Lagrange, Observatoire de la C\^ote d’Azur, CNRS, Universit\'e C\^ote d’Azur, Nice, France\\
\inst{*}\email{weisserd@mcmaster.ca}
}

  \abstract 
   {Tracing the compositional link between terrestrial super-Earths and their host stars provides clues to their dominant formation pathway. By constraining the stellar abundances of refractory elements, we can predict the core mass fractions (CMFs) of their super-Earths. The level of agreement between this prediction and the planetary CMF derived from their masses and radii can reveal past formation processes, like mantle stripping and water-rich formation plus sequestration in the planet’s core. Here, we present the first results from the Near Infrared Planet Searcher (NIRPS) GTO CMF subprogram: an intensive radial velocity campaign to refine masses and compute host stellar abundances of three hot super-Earths around M dwarfs (GJ\,1132\,b, GJ\,1252\,b, and LTT\,3780\,b), calculating masses of $1.69 \pm 0.15 M_\oplus$, $1.54 \pm 0.18 M_\oplus$, and $2.34 \pm 0.10 M_\oplus$ respectively. We measure the CMFs of these and {six} further hot super-Earths with precise masses already available in the literature to 10--15\% precision. We compare these to CMF predictions made from measuring the Fe, Mg, and Si abundances of their host stars measured from the NIRPS spectra. We find that the CMFs of these planets are smaller than expected from their host stellar abundances, to a statistically significant degree. This discrepancy is suggestive of significant reservoirs of water, and while these planets are too hot to harbor surface water, they likely have interior water mass fractions of $\sim 1\%$.}
   \maketitle

\section{Introduction} \label{sec:intro}

Characterizing the composition of exoplanets is crucial for understanding the pathways that drive their formation. Planet formation models predict qualitatively different distributions of small planet compositions between M dwarfs and Sunlike stars, finding that low-mass, water-rich planets are expected to be common around M dwarfs while appearing with a lower frequency around FGK stars \citep{MiguelCridland2020,BurnSchlecker2021,VenturiniRonco2024}. Some observational evidence, too, supports different planet formation mechanisms around M dwarfs compared to Sunlike stars. For example, the slope of the radius valley around early M dwarfs is markedly shallower \citep{CloutierMenou2020,GaidosAli2024,BonfantiBrady2024,HoRogers2024,WanderleyCunha2025}, which points towards the emergence of the M dwarf radius valley as a consequence of planet formation processes rather than being due to thermally-driven mass loss. In this paradigm, sub-Neptunes are predominantly water-rich worlds, rather than having H$_2$/He-dominated atmospheres. Additional supporting empirical evidence for a population of water worlds around M dwarfs comes from the detailed characterization of individual planetary systems with masses or atmospheres suggestive of a water-rich composition \citep[e.g.,][]{DiamondLoweKreidberg2022,PiauletBenneke2023,CadieuxPlotnykov2024,CoulombeBenneke2025} and from planet population-level studies based on mass and radius characterization \citep[e.g.,][]{LuquePalle2022,CherubimCloutier2023,HoRogers2024,ParcBouchy2024}. Empirically assessing the bulk water content of small planets around M dwarfs, however, remains an open problem that needs to be addressed to properly compare to formation model predictions.

Small close-in planets, with their high equilibrium temperatures ($T_{\rm eq} \gtrsim 800$ K), are not expected to have a significant condensed surface water layer or a thick atmosphere, as surface water would be vaporized and any atmospheric gas would undergo rapid thermal escape \citep{Lopez2017,OwenWu2017}. As such, it is commonly assumed that the interiors of such planets are composed of two components: an iron-dominated core that is differentiated beneath a primarily silicate mantle. Assuming that planets are composed of only two compositional layers avoids the model degeneracy introduced when attempting to infer more than two compositional mass fractions from the planet's measured mass and radius alone. The compositions of these terrestrial planets are well-described by a single parameter detailing the relative masses of the rocky mantle and iron core: the core mass fraction (CMF). The CMF is defined as the fraction of the planet's total mass contained within its (assumed) differentiated iron core. However, precisely measuring planetary CMF values requires very precisely constrained masses and radii. Even typical measurement precisions of $20\%$ on planet mass and $5\%$ on planet radius result in large CMF uncertainties approaching $20\%$ \citep{otegi2020impact,PlotnykovValencia2024}. Ultra-precise masses and radii (to $\sim10\%$ and $\sim3\%$ precision, respectively) are needed to recover CMF uncertainties that are useful to study the compositional connection between planets and their host stars, and to constrain planetary formation and evolutionary processes.

Oftentimes, due to the uncertainties in planetary masses or radii, one may rely on the refractory abundances of a planet's host star to set a prior on its CMF. Planets form from the same primordial cloud of material as their host stars such that it is often assumed that the relative chemical abundances of nonvolatiles like iron, magnesium, and silicon are common between the planet and star. Many studies have used stellar refractory abundances as proxies on planetary compositions \citep[e.g.,][]{dorn2015can, DornHinkel2017,DornVenturini2017,unterborn2018inward}. As data quality improved, it became possible to test the strength of this correlation directly \citep{SantosAdibekyan2015,SantosAdibekyan2017,PlotnykovValencia2020,AdibekyanDorn2021,SchulzeWang2021,UnterbornDesch2023,BrinkmanPolanski2024,BrinkmanWeiss2025,BehmardBrinkman2025}. However, these studies have focused largely on planets around FGK stars rather than planets around M dwarfs. In addition, these studies have often struggled to draw conclusions for individual planetary systems due to their uncertainties involved, particularly in planetary mass. As a result, they often study planets on a demographics level instead, which have yielded a variety of results that differ substantially in the strength of the star-planet compositional connection: some studies have found very little evidence for a star-planet compositional connection \citep[e.g.,][]{BrinkmanPolanski2024}, while others have found a very strong correlation between planetary and host stellar composition \citep[e.g.,][]{AdibekyanDorn2021}.

Small planets around M dwarfs provide our best chances of measuring a planet's mass and radius to the precision necessary for composition analysis. Hot super-Earths have small radii and low masses that make them generally difficult to study. Fortunately, small, low-mass M dwarfs produce larger planetary signals in transit and radial velocity (RV) observations, allowing for the precise characterization of these planets' masses and radii. Conversely, accurate stellar abundances of M dwarfs are difficult to measure. Atomic lines are often hidden or blended with deep molecular bands, which suppress the continuum in many wavelength regimes, particularly in the optical and the near-infrared (NIR) where M dwarfs are brightest \citep[e.g.,][]{SoutoCunha2017,SarmentoRojasAyala2021,JahandarDoyon2024}. This complicates the recovery of accurate stellar abundances for M dwarfs. As such, previous studies comparing planetary and stellar compositions have focused largely on hotter stars. Fortunately, recent advancements in NIR spectrographs (such as the Infrared Doppler Instrument \citep[IRD;][]{TamuraSoto2012}, the Habitable-zone Planet Finder
Spectrograph \citep[HPF;][]{MahadevanRamsey2014}, M-dwarf Advanced Radial velocity Observer Of Neighboring eXoplanets \citep[MAROON-X][]{SeifahrtBean2018}, SpectroPolarimètre InfraRouge \citep[SPIRou;][]{DonatiKouach2020}, and the Near Infrared Planet Searcher \citep[NIRPS;][]{BouchyDoyon2025}) have allowed for improvements in extremely precise mass determinations of these exoplanets while enabling stellar abundance analyses from high-resolution spectroscopy \citep[e.g.,][]{HejaziCrossfield2023,JahandarDoyon2025} that opens new perspectives for the derivation of precise characterization of M dwarf stellar abundances, allowing for direct comparison between planetary and stellar compositions.

In this paper, we present the first results of the Near Infrared Planet Searcher \citep[NIRPS;][]{BouchyDoyon2025} GTO CMF subprogram to measure precise planetary masses and stellar abundances of {nine} hot super-Earths orbiting M dwarfs. We measure the planetary CMFs and compare those values to the CMFs predicted from the host stars' refractory abundances to search for evidence of past formation or evolutionary processes. We present intensive RV follow-up observations with NIRPS and HARPS of three planets to improve their mass measurement precisions, and we additionally measure stellar refractory abundances for these plus five other hot super-Earth hosts. Our paper is structured as follows: In Section \ref{sec:obs}, we discuss our target selection and the data used for our RV and abundance analyses. In Section \ref{sec:stellchar}, we discuss how we determine stellar parameters and abundances, and report measured abundances for Mg, Si, and Fe (and Ti in some cases). In Section \ref{sec:rv}, we discuss our RV model used to determine planetary masses, and in Section \ref{sec:cmf}, we derive the CMFs of our planets, both from the planetary masses and radii and from stellar abundances. Lastly, in Section \ref{sec:disc}, we compare our results and discuss their implications for the compositions of rocky planets around M dwarfs. We conclude with a summary of our findings in Section \ref{sec:conclusions}.

\section{Observations} \label{sec:obs}
All targets in this sample were observed with NIRPS \citep{BouchyDoyon2025}. NIRPS is a NIR echelle spectrograph designed for precision radial velocities, mounted alongside the High Accuracy Radial Velocity Searcher \citep[HARPS;][]{MayorPepe2003} optical spectrograph on the ESO 3.6-m telescope at the La Silla Observatory in Chile. NIRPS observes in the $YJH$ bands ($980-1800$ nm), can be used simultaneously with HARPS, and has two observing modes: a High Accuracy mode (HA, $R \approx 88\,000, 0.4^{\prime\prime}$ fiber) and a High Efficiency mode (HE, $R \approx 75\,000, 0.9^{\prime\prime}$ fiber). All HARPS observations taken simultaneously with NIRPS used for RV analysis in this paper were taken in HE mode.

The majority of our observations were made through the NIRPS Guaranteed Time Observations (GTO) program, as part of the GTO's second Work Package (WP2) dedicated to mass and density measurements of transiting exoplanets around M dwarfs \citep{ArtigauBouchy2024}. WP2 is composed of several subprograms, including the CMF subprogram, whose aim is to obtain precise masses of close-in transiting super-Earths with the intent of characterizing these planets' core mass fractions and constraining their composition. 

We obtain observations of {nine} hot transiting super-Earths across eight systems amenable to estimating CMFs. Three of these targets -- GJ\,1132\,b, GJ\,1252\,b, and LTT\,3780 b -- were selected for intensive RV follow-up as part of the CMF subprogram for the purpose of obtaining $10\sigma$ precision mass determinations, which is the level of precision identified as necessary to sufficiently constrain their compositions (giving core mass fraction uncertainties on the order of $10-15\%$; \citealp{PlotnykovValencia2024}). These targets were selected for intensive RV follow-up because they are the planets for which $10\sigma$ masses can be most efficiently obtained with NIRPS when combined with literature RVs \citep[calculated using the formalism in][]{CloutierDoyon2018}.

We select the other {six} targets in our sample -- GJ\,357\,b, HD\,260655\,b, L\,98-59\,b, {L\,98-59\,c,} LHS\,1140\,c, and TOI-270\,b -- by selecting all M dwarfs ($T_\textrm{eff} < 3900 \textrm{ K}$) with small ($R_p < 1.5 R_\oplus$), hot ($T_\textrm{eq} > 373 \textrm{ K}$) planets with precisely determined ($>5\sigma$) masses that either had existing or planned NIRPS observations. These targets were identified as very favorable for obtaining stellar refractory abundances from NIRPS observations, which we can use to compare the planetary composition to expectations from the host star (discussed in more detail in Section \ref{subsec:cmf-stellar}). All {six} of these target planets have reported radii and mass measurements above $\gtrsim 5\sigma$ precision, which are expected to produce CMF uncertainties of roughly $15-20\%$ \citep{PlotnykovValencia2024}. While obtaining ultra-precise masses for these targets would be prohibitively expensive with NIRPS, they remain viable targets for measuring CMFs using literature masses and comparing to stellar abundances, and can still be used to investigate the compositional connection between the planets and their host stars.

\subsection{GJ\,1132} \label{subsec:obs-gj1132}

GJ\,1132\,b, first discovered by \citet{Berta-ThompsonIrwin2015} using ground-based photometry from the MEarth-South Observatory \citep{IrwinBertaThompson2015}, is a rocky planet {transiting} an M4.5V star. More recently, \citet{XueBean2024} analyzed photometric data of this system from the Transiting Exoplanet Survey Satellite (TESS) and the James Webb Space Telescope (JWST), calculating the orbital period of this planet to be $1.62892911 \pm 3 \times 10^{-6}$ days and the radius of the planet relative to its star $R_p / R_\star$ to be $0.04943 \pm 0.00015$. {GJ 1132 is a multiplanet system, with a second planetary signal at $8.93$ days \citep{BonfilsAlmenara2018}; however, as GJ 1132\,c does not transit, its radius is unknown and thus its composition cannot be effectively constrained.}

We use 128 archival observations of GJ\,1132 from HARPS taken between June 6, 2016 and June 21, 2017, with a median uncertainty of $2.4$ m/s and a mean dispersion of $4.8$ m/s. These RVs are used for and described in more detail in \citet{BonfilsAlmenara2018}, a previous study on the system. As these radial velocities were extracted using the template-matching method, which is expected to produce similar results as the line-by-line method \citep{ArtigauCadieux2022}, we did not rereduce or extract these radial velocities; the final RV measurements were taken directly from \citet{BonfilsAlmenara2018}.

We additionally acquire 160 NIRPS spectra and 73 HARPS spectra over 80 nights, all with 900\,s exposures, taken between April 4, 2023 and July 4, 2024.

The NIRPS data were reduced and automatically corrected for telluric contamination using version 0.7.292 of APERO \citep{CookArtigau2022}, while the HARPS data were reduced with the ESPRESSO DRS 3.2.5 \citep{PepeCristiani2021}, adapted for use with HARPS and available through the Data \& Analysis Center for Exoplanets (DACE) platform. The RV extraction was performed using the line-by-line (LBL) method \footnote{\url{https://lbl.exoplanets.ca/}}, using version 0.65.002 of the $texttt{LBL}$ package. After extraction (which bins observations on a nightly basis and removes observations with peak $H$-band S/Ns below $10$), we end up with 81 NIRPS RVs with a median uncertainty of $2.4$ m/s and a root-mean-square (rms) dispersion of $5.3$ m/s, and 57 HARPS RVs with a median uncertainty of $3.8$ m/s and an rms dispersion of $6.4$ m/s.

We additionally use the NIRPS observations taken here for the purposes of stellar abundance analysis. These NIRPS observations have a median signal-to-noise (S/N) in the $H$ band of $76.0$.

\subsection{GJ\,1252} \label{subsec:obs-gj1252}

GJ\,1252\,b, first discovered by \citet{ShporerCollins2020}, is a rocky planet {transiting} an M3V star. Most recently, \citet{KokoriTsiaras2023} analyzed photometric data of this system from TESS, calculating the orbital period of this planet to be $0.5182464 \pm 2.3\times 10^{-6}$ days and the radius of the planet relative to its star $R_p / R_\star$ to be $0.02802 \pm 0.00090$. {GJ 1252 has an additional planet candidate in the system, GJ 1252\,(c), detected through radial velocities at a period of $17.5$ days \citep{LuquePalle2022}; however, as this planet candidate does not transit, its radius is unknown and thus its composition cannot be effectively constrained.}

We use 48 archival observations of GJ\,1252 from HARPS taken between September 19, 2019 and November 18, 2019, with a median uncertainty of $2.0$ m/s and a dispersion of $4.9$ m/s. These radial velocities are analyzed in \citet{ShporerCollins2020} and \citet{LuquePalle2022}, two previous studies on the system, and are described in more detail in \citet{LuquePalle2022}. Similarly to the archival HARPS RVs of GJ\,1132, we adopt the template-matching RVs directly from \citet{LuquePalle2022} and do not perform an LBL extraction.

We additionally acquire 171 NIRPS spectra and 148 HARPS spectra over 87 nights, all with 900s exposures, obtained between April 12, 2023 and November 26, 2024. All of these RVs were reduced and extracted in nearly the same manner as described in Section \ref{subsec:obs-gj1132}. Due to heavy telluric contamination at certain epochs, we instead reduce NIRPS observations with version 3.2.0 of the NIRPS DRS \citep{BouchyDoyon2025} with a more aggressive masking of pixels (discussed in more detail in Srivastava et al. in prep). This resulted in 85 NIRPS RVs with a median uncertainty of $2.1$ m/s and a dispersion of $4.2$ m/s and 65 HARPS RVs with a median uncertainty of $2.0$ m/s and a dispersion of $8.3$ m/s.

We additionally use the NIRPS observations taken here for the purposes of stellar abundance analysis. These NIRPS observations have a median S/N in the $H$ band of $103.4$.

\subsection{LTT\,3780} \label{subsec:obs-ltt3780}

LTT\,3780\,b, co-discovered by \citet{CloutierEastman2020} and \citet{NowakLuque2020}, is a rocky planet {transiting} an M3.5V star. Most recently, \citet{BonfantiBrady2024} analyzed photometric data of this system from the TESS and The CHaracterising ExOPlanet Satellite \citep[CHEOPS;][]{BenzBroeg2021}, calculating the orbital period of this planet to be $0.76837931_{-0.00000042}^{+0.00000039}$ days and the radius of the planet relative to its star $R_p / R_\star$ to be $0.03212_{-0.00072}^{+0.00068}$. {LTT 3780 is a multiplanet system, with a second transiting planet, LTT\,3780\,c, orbiting at a period of $12.25$ days \citep{BonfantiBrady2024}; however, as it has a radius of $2.39 R_\oplus$, it is likely not terrestrial and so is not a target for this analysis.}

Despite LTT\,3780\,b being identified as a target for intensive RV follow-up through the CMF subprogram, we only obtain 17 nights of observations, as during the course of our observations, a new study was published on the system, \citet{BonfantiBrady2024}. This paper published RVs from several spectrographs and obtained a mass precision for LTT\,3780\,b beyond our $10\sigma$ threshold. We ceased observations on this target with NIRPS soon after as a result, though we rederive the mass of LTT\,3780\,b nonetheless.

We use radial velocity data measurements of LTT\,3780 from four spectrographs. 19 exposures of 1200 seconds are obtained from MAROON-X in the red and blue channels each between February 22, 2021 and June 4, 2021, described in more detail in \citet{BonfantiBrady2024}, with median uncertainties of $0.5$ m/s for the red channel and $1.0$ m/s for the blue channel, and dispersions of $5.2$ m/s for the red channel and $4.2$ m/s for the blue channel. The red and blue channels are treated as separate instruments for the purpose of the radial velocity analysis performed in this paper \citep[e.g.,][]{TrifonovCaballero2021}.

52 exposures are obtained from CARMENES between December 27, 2019 and February 19, 2020, described in more detail in \citet{NowakLuque2020}, with a median uncertainty of $1.6$ m/s and a dispersion of $2.6$ m/s.

33 exposures of 2400 seconds are obtained from HARPS between June 21, 2019 and February 24, 2020, described in more detail in \citet{CloutierEastman2020}, with a median uncertainty of $1.3$ m/s and a dispersion of $4.7$ m/s. 30 exposures of 1800 seconds are obtained with the HARPS-N spectrograph \citep{ConsentinoLovis2012} between December 14, 2019 and March 15, 2020, described in more detail in the same paper, with a median uncertainty of $1.4$ m/s and a dispersion of $9.5$ m/s.

In addition, we use 37 NIRPS spectra and 22 HARPS spectra over 17 nights, all with 900s exposures, obtained between April 1, 2023 and November 27, 2023. All of these RVs were reduced and extracted in the same manner as described in Section \ref{subsec:obs-gj1132}, resulting in 17 NIRPS RVs with a median uncertainty of $1.6$ m/s and a dispersion of $3.7$ m/s. The HARPS RVs were erroneously taken over two different observing modes, the high-efficiency mode EGGS and the high-accuracy mode HAM. As these were taken in two different observing modes, they are reduced and extracted as if they are separate instruments. This results in 11 HAM HARPS RVs with a median uncertainty of $4.8$ m/s and a dispersion of $29.0$ m/s, and 11 EGGS HARPS RVs with a median uncertainty of $1.9$ m/s and a dispersion of $3.8$ m/s. However, as the expected information content of the radial velocities from the HARPS observations (in either the HAM or EGGS mode) simultaneous with NIRPS is negligible compared to the other radial velocities, we elect not to use the HARPS RVs measured simultaneously with NIRPS in this radial velocity analysis.

We additionally use the NIRPS observations taken here for the purposes of stellar abundance analysis. These NIRPS observations have a median S/N in the H band of $105.7$.

{For the rest of the targets discussed here, we use these spectra for the sole purpose of measuring elemental abundances and not for mass measurement improvements.}

\subsection{GJ\,357} \label{obs-gj357}

GJ\,357\,b, first discovered by \citet{LuquePalle2019}, is a hot  rocky planet {transiting} the M2.5V star GJ\,357. This discovery paper measured the mass of GJ\,357\,b to be $1.84 \pm 0.31 M_\oplus$ $(5.9\sigma)$. Most recently, \citet{OddoDragomir2023} analyzed photometric data of this system from TESS and CHEOPS, calculating the orbital period of this planet to be $3.930600 \pm 0.000002$ days and the radius of the planet relative to its star $R_p / R_\star$ to be $0.0309 \pm 0.0010$. {GJ\,357 host two other confirmed nontransiting planets, GJ\,357\,c and GJ\,357\,d \citep{LuquePalle2019}; however, as they are nontransiting, their radii are unknown and their compositions thus cannot be effectively constrained.}

4 NIRPS observations over 2 nights, with 600s exposures, were taken on March 1, 2025 and March 4, 2025. These NIRPS observations have a median S/N in the $H$ band of $199.7$.

\subsection{HD\,260655} \label{obs-hd260655}

HD\,260655\,b, first discovered by \citet{LuqueFulton2022}, is a  hot rocky super-Earth {transiting} the M0V star HD\,260655. This discovery paper measured the mass of HD\,260655\,b to be $2.14 \pm 0.34 M_\oplus$ $(6.3\sigma)$ using radial velocities from the high-resolution spectrographs HIRES and CARMENES. \citet{LuqueFulton2022} analyzed photometric data of this system from TESS, calculating the period of HD\,260655\,b to be $2.76953 \pm 0.00003$ days and the radius of the planet relative to its star $R_p / R_\star$ to be $0.02586 \pm 0.00046$. {HD\,260655 is known to host another transiting planet, HD\,260655\,c \citep{LuqueFulton2022}. However, this planet has a radius of $1.53 R_\oplus$ and a relatively low density of $4.7$ g/cm$^3$, meaning it is less dense than a planet composed of entirely MgSiO$_3$ would be at its mass. As such, we cannot be confident this planet is entirely terrestrial, and so do not select it as a target for this analysis.}

We obtained 106 NIRPS observations over 2 nights, with a mean exposure time of 122s, were taken on November 29, 2023 and February 2, 2024, as part of the COMPASS subprogram in NIRPS' Work Package 3 for the purposes of transmission spectroscopy. These NIRPS observations have a median S/N in the H band of $123.5$.

\subsection{L 98-59} \label{obs-l9859}

L\,98-59\,b, first discovered by \citet{KostovSchlieder2019}, is a hot rocky planet {transiting} the M3V star L\,98-59. Most recently, \citet{CadieuxLHeureux2025} measured the mass of L\,98-59\,b to be $0.46 \pm 0.10 M_\oplus$ $(4.6\sigma)$ by analyzing RVs from HARPS and the Echelle SPectrograph for Rocky Exoplanets and Stable Spectroscopic Observations \citep[ESPRESSO;][]{PepeCristiani2021} (first obtained in \citealp{CloutierAstudillo-Defru2019} and \citealp{DemangeonZapateroOsorio2021}) alongside transit timing variations from TESS and JWST observations. \citet{CadieuxLHeureux2025} also analyzed photometric data of this system from TESS, calculating the orbital period of this planet to be $2.253114 \pm 0.0000004$ days and the radius of the planet relative to its star $R_p / R_\star$ to be $0.0243 \pm 0.0003$.

{L\,98-59\,c is a hot planet transiting L 98-59 as well. Similarly discovered by \citet{KostovSchlieder2019}, \citet{CadieuxLHeureux2025} recently measured the mass of L\,98-59\,c to be $2.00\pm 0.13 M_\oplus$ $(15\sigma)$, and the orbital period and radius of L\,98-59\,c relative to its star to be $3.6906764 \pm 0.0000004$ days and $0.0386 \pm 0.0004$, respectively, using the same data. L 98-59 is known to host three other confirmed planets, one of which, L 98-59 d, is transiting \citep{KostovSchlieder2019,CadieuxLHeureux2025}. However, L 98-59 d has a radius above $1.6 R_\oplus$ and thus is likely not terrestrial, so it is not a target for our analysis.}

301 NIRPS observations over 5 nights, with exposure times between 100s and 300s, were taken between April 11, 2023 and March 9, 2024 as part of the COMPASS subprogram in the NIRPS-GTO WP3, studying systems of mutually misaligned small planets, for the purposes of characterizing system architectures and performing transmission spectroscopy. These NIRPS observations have a median signal-to-noise S/N in the H band of $58.3$.

\subsection{LHS\,1140} \label{obs-lhs1140}

LHS\,1140 c, first discovered by \citet{DittmannIrwin2017}, is a hot rocky planet {transiting} the M4.5V star LHS\,1140. Most recently, \citet{CadieuxPlotnykov2024} measured the mass of LHS\,1140 c to be $1.91 \pm 0.06 M_\oplus$ $(31.8\sigma)$ using RV measurements from ESPRESSO. \citet{CadieuxPlotnykov2024} analyzed photometric data of this system from TESS and the Spitzer Space Telescope \citep{WernerRoellig2004}, calculating the orbital period of this planet to be $3.777940 \pm 0.000002$ days. \citet{CadieuxDoyon2024} also analyzed photometric data of this system from JWST, calculating the radius of the planet relative to its star $R_p / R_\star$ to be $0.05312 \pm 0.00028$. {LHS\,1140 is known to host another transiting planet, LHS\,1140\,b; however, this planet is unlikely to be a rocky super-Earth due to its low bulk density and radius of $1.73 R_\oplus$ \citep{CadieuxPlotnykov2024}, and so is not used for this analysis.}

29 NIRPS observations over 9 nights, with 600s exposures, were taken between November 27, 2022 and December 5, 2022 as part of NIRPS commissioning. These NIRPS observations have a median S/N in the H band of $56.1$.

\subsection{TOI-270} \label{subsec:obs-toi270}

TOI-270\,b, first discovered by \citet{GuntherPozuelos2019}, is a hot rocky super-Earth {transiting} the M3V star TOI-270. Most recently, \citet{KayeVissapragada2022} measured the mass of TOI-270\,b to be $1.48 \pm 0.18M_\oplus$ $(8.2\sigma)$ using a combination of transit timing variations derived using data from TESS and radial velocities obtained from ESPRESSO. In addition, \citet{VanEylenAstudillo-Defru2021} also analyzed photometric data of this system from TESS, calculating the orbital period of this planet to be $3.3601538 \pm 0.0000048$ days, while analysis of JWST data from \citet{CoulombeBenneke2025} found the radius of the planet relative to its star $R_p / R_\star$ to be $0.03144 \pm 0.00010$. {TOI-270 hosts two other transiting planets, TOI-270\,c and TOI-270\,d \citep{VanEylenAstudillo-Defru2021}; however, as both planets have very large radii above $2 R_\oplus$, they are likely not terrestrial and so are not used as targets for this analysis.}

In addition, 239 NIRPS observations over 8 nights, with 400s exposures, were taken between November 20, 2023 and December 26, 2024 as part of the COMPASS subprogram in the NIRPS-GTO WP3, studying systems of mutually misaligned small planets, for the purposes of characterizing system architectures and performing transmission spectroscopy. These NIRPS observations have a median S/N in the H band of $62.9$.

\section{Stellar characterization} \label{sec:stellchar}

\subsection{Stellar parameters} \label{subsec:stellchar-params}

We obtain the masses of our M dwarf targets from the relations reported by \citet{MannDupuy2019}. This study contains empirical relations between the mass of M dwarfs and their $K$-band absolute magnitudes (specifically, we use the relation in Equation 4 of that work, using the $n = 5$ coefficients given in Table 6 of that work). We calculate the $K$-band absolute magnitudes used in that relation from their apparent magnitudes, obtained from the Two Micron All Sky Survey \citep[2MASS;][]{SkrutskieCutri2006} catalog of point sources, and their parallaxes, obtained from Gaia Data Release 3 \citep{GaiaCollaborationVallenari2023}.

We obtain the radii of our M dwarf targets from the relations in \citet{MannFeiden2015}. This paper contains empirical relations between radii of M dwarfs and their $K$-band absolute magnitudes (specifically, we use the relation in Equation 4 of that work, using the coefficients given in Table 1 of that work). We use the same $K$-band absolute magnitudes calculated for our target mass relations. With these masses and radii, $\log g$  values for all of our target stars are derived.

While we measure stellar metallicities through observations of their spectra (described in Section \ref{subsec:stellchar-abund}), we nonetheless need initial guesses of stellar effective temperature and metallicity with which to constrain our stellar abundance calculations. We obtain these initial guesses from previous studies characterizing these host stars. {The effective temperatures used are provided in Table \ref{table:stellar-params}. However, to distinguish between the stellar metallicities we calculate in Section \ref{subsec:stellchar-abund} and the initial metallicity guesses, we present the latter here.}
For GJ\,1132, {we use} $-0.17 \pm 0.15$ \citep{XueBean2024}; for GJ\,1252, {we use} $0.10 \pm 0.10$ \citep{CrossfieldMalik2022}; for LTT\,3780, {we use} $0.06 \pm 0.11$ \citep{BonfantiBrady2024}; for GJ\,357, $-0.12 \pm 0.16$ \citep{SchweitzerPassegger2019}; for HD\,260655, $-0.43 \pm 0.10$ \citep{MarfilTabernero2021}; for L 98-59, $-0.46 \pm 0.26$ \citep{DemangeonZapateroOsorio2021}; for LHS\,1140, $-0.15 \pm 0.09$ \citep{CadieuxPlotnykov2024}; for TOI-270, $-0.20 \pm 0.12$ \citep{VanEylenAstudillo-Defru2021}. Note that the effective temperature and metallicity originally reported for HD\,260655 in \citet{MarfilTabernero2021} are substantially more precise than used here, at $3803 \pm 10$\,K and $-0.43 \pm 0.04$ respectively; however, it has been found that varying methods for estimating M dwarf stellar parameters have differences well in excess of these uncertainties \citep{PasseggerBello-Garcia2022}, and so we choose to inflate these uncertainties to $50$\,K and $0.1$ dex for use in this paper.

{All eight host stars in our sample have very long rotation periods: HD 260655 has the shortest rotation period in our sample at $37.5$ days \citep{LuqueFulton2022}, and LHS 1140 has the longest at $131$ days \citep{CadieuxPlotnykov2024}. As all stars in our sample are slow rotators, they are not expected to significantly affect any estimated stellar parameters (although stellar activity sources such as sunspots may still induce signals in radial velocity time series, which we model in Section \ref{sec:rv}).}

\subsection{Stellar elemental abundances} \label{subsec:stellchar-abund}

We measure stellar elemental abundances of our target stars based on the spectral synthesis of individual lines in high-resolution near-IR spectra (here using NIRPS), as has been demonstrated in the literature \citep[e.g.,][]{SoutoCunha2017,HejaziCrossfield2023}. Here we provide an overview of the methodology described by \cite{Gromek2025} and Gromek et al. (in prep.) to measure the abundances (and associated uncertainties) of the refractory elements Mg, Si, Fe, and Ti for select cool stars.

We perform our abundance analyses on the order-merged, telluric-corrected NIRPS spectral template produced by the \texttt{APERO} pipeline. We use the \texttt{iSpec} spectral analysis tool \citep{Blanco-CuaresmaSoubiran2014,Blanco-Cuaresma2019} to perform a global continuum normalization using a spline fit, apply an RV correction by cross-correlating the spectrum with rest wavelength line positions from the VALD linelist \citep{KupkaDubernet2011}, and identify prominent atomic and OH absorption features in our NIRPS spectra whose absorption depths exceed 5\%. The line identification step provides us with our initial line list for our abundance analysis (including identified wavelength boundaries of each line, denoting a "line region"), which we subsequently refine using visual inspection to remove contaminated features.

\begin{figure}
    \centering
    \includegraphics[width=1.1\linewidth]{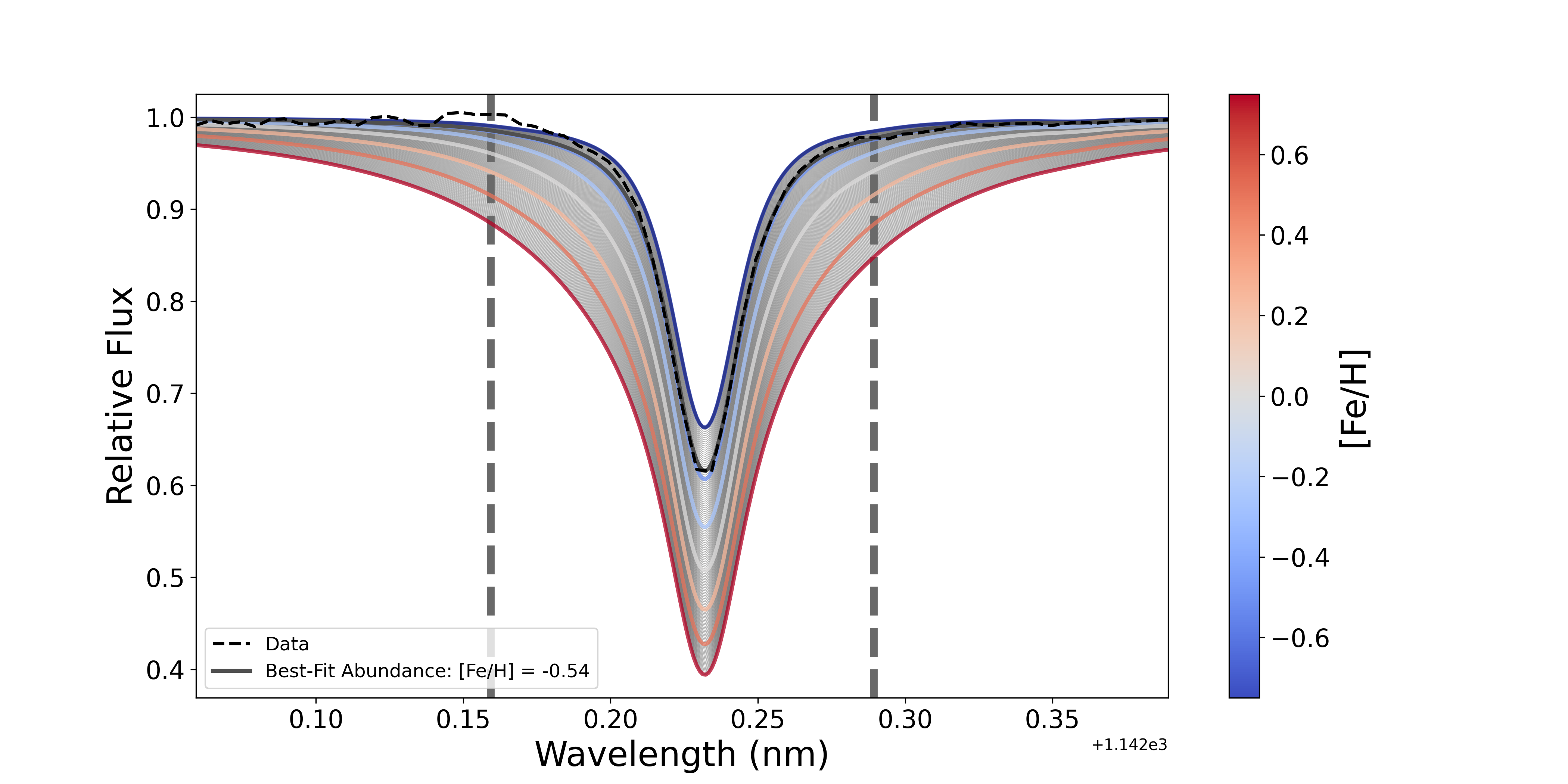}
    \caption{The model fit to the $1142.2$ nm refractory Fe I line for the star HD\,260655. Colored lines represent different synthetic spectra, while the fainter gray lines represent the interpolated grid of spectra being fit to the data, shown as a black dashed line. The best-fit abundance of $\textrm{[Fe/H]} = -0.54$ is shown as a darker gray line, tracking very closely with the data over the line region, delineated on either side by dashed gray lines.}
    \label{fig:fe-line-sample}
\end{figure}

We use the Turbospectrum radiative transfer code \citep{Plez2012} and the MARCS model atmospheres \citep{GustafssonEdvardsson2008} to calculate our synthetic spectra. For each spectral line, we fix the stellar effective temperature $T_\textrm{eff}$, surface gravity $\log\textrm{ g}$, metallicity [M/H], macroturbulence $v_\textrm{mac}$, and microturbulence $v_\textrm{mic}$, and generate seven synthetic spectra over the surrounding line region by varying the abundance of the element in question [X/H] across an evenly spaced grid from $[-0.75, +0.75]$ dex in increments of $0.25$ dex. Some lines in the NIRPS spectrum show a significant flux offset with the synthetic spectra due to a poor continuum normalization near the line region. To ensure a better fit between the model and the data, we identify through manual inspection lines for which a further pseudo-continuum normalization step would significantly improve the fit between the model and data, and for these lines we fit for the constant flux offset for each line that would minimize the $\chi^2$ between the model and data in the region of the continuum within $5$ Å of the line region. We then interpolate all of the spectra to create a much finer grid of spectra with a spacing of $0.01$ dex. Finally, we identify the elemental abundance that minimizes the $\chi^2$ statistic between the spectral model and the data over the line region. We show an example of a model fit to a single line in Figure \ref{fig:fe-line-sample}.

Our initial abundance calculations assume fixed stellar parameters. We use $T_\textrm{eff}$ and [M/H] measurements from the literature, as discussed in Section \ref{subsec:stellchar-params}. We adopt $\log\textrm{ g}$ values calculated from our derived stellar masses and radii. $v_\textrm{mac}$ and $v_\textrm{mic}$ are nuisance parameters that cannot be reliably measured from our NIRPS spectra. As such, we use the OH I lines in the spectrum to constrain the stellar $v_\textrm{mac}$ and $v_\textrm{mic}$, as these lines are prominent, numerous, and highly sensitive to changes in these parameters \citep{SoutoCunha2017,HejaziCrossfield2023}. We calculate the $\chi^2$ of the best-fit oxygen abundance [O/H] across a  wide grid of $v_\textrm{mac}$ and $v_\textrm{mic}$, parameters, fitting a Gaussian distribution to the logarithm of the median $\chi^2$ (normalized to the best-fit $\chi^2$ for each line) to calculate our best-fit $v_\textrm{mac}$ and $v_\textrm{mic}$ and estimate their corresponding uncertainties. In some cases, most commonly with $v_\textrm{mic}$, fitting a Gaussian distribution gives unreasonably wide uncertainties or physically implausible parameter ranges; in those cases, we identify a plausible range of parameters by eye and use the center of this range as our final turbulence parameter instead.

With these parameters now known, we calculate the best-fit elemental abundance for all spectral lines in our spectra corresponding to our elements of interest, and calculate the stellar abundance of a given refractory element to be the weighted average of the abundances of these spectral lines, where the weight of each line is equal to the maximum depth of the spectral line divided by the root-mean-square error of the best-fit model across the line region.

We proceed with quantifying the uncertainties on our abundance measurements by considering the random errors across multiple lines of the same element and systematic errors introduced by uncertainties in the input stellar parameters. We calculate the random error of a given abundance $\sigma_\textrm{rand}$ as the weighted standard deviation of the mean abundance of all lines of a corresponding element, with the same weights as the ones used to compute the stellar abundance of the corresponding element. We additionally calculate the errors caused by the uncertainties in $T_\textrm{eff}$, [M/H], $v_\textrm{mic}$, and $v_\textrm{mac}$ (denoted $\sigma_\textrm{Teff}$, $\sigma_\textrm{[M/H]}$, $\sigma_\textrm{vmic}$, and $\sigma_\textrm{mac}$) by resampling each of these parameters ten times while keeping the other parameters fixed, repeating our abundance measurements with each new realization and defining the uncertainty from that parameter as the standard deviation in our final abundances across these resamples. As our uncertainties in $\log\textrm{ g}$ are so small, and it has such a small impact on the synthetic spectra calculations, the error caused by our uncertainty in $\log\textrm{ g}$ would be negligible, and so we do not calculate it here. We then add each error contributor in quadrature to calculate our final uncertainties in elemental abundance. The final uncertainties for each source, as well as our final abundances, are given in Table \ref{table:abundances-appendix}.

We use this process to calculate the refractory abundances [Fe/H], [Mg/H], and [Si/H] for all eight host stars in our sample. We then calculate refractory ratios Fe/Mg and Mg/Si, as these ratios largely control the internal structure models of terrestrial planets, including their CMFs. We also calculate the metallicity proxy [M/H] based on Mg, Si, and Fe only for each of these stars, using the prescription given by \citet{HinkelYoung2022}. Our final refractory abundances, as well as our refractory ratios and resulting metallicities, are given in Table \ref{table:stellar-params}.

We note that for the two coolest stars in our sample, GJ\,1132 and LHS\,1140, the [Mg/H] and [Si/H] results may be unreliable. For these cool stars, there are very few Mg and Si lines, and the quality of both the spectral lines and the model fits to these lines are relatively poor. In addition, as Mg and Si are both alpha elements, we would expect the ratio Mg/Si to be relatively similar across all stars, but the Mg/Si values calculated from these results are thus very discrepant from solar. As such, for these two stars, we additionally calculate [Ti/H] to serve as a proxy for [Mg/H] and [Si/H]; Ti is also an alpha element, so we would expect [Ti/H] to be very similar to [Mg/H] and [Si/H], but there are many more Ti lines in our sample and the model fits are substantially improved for Ti. Because of this, while we report the [Mg/H] and [Si/H] values we find in Tables \ref{table:stellar-params} and \ref{table:abundances-appendix}, for the purpose of our interior structure modeling in Section \ref{sec:cmf}, we assume that the [Mg/H] and [Si/H] values are equal to the more reliable [Ti/H] values calculated for GJ 1132 and LHS\,1140. (While HD\,260655 does have an Mg/Si value discrepant from solar, its spectra contains many more Mg and Si lines, and both the quality of these lines and the model fits to them are substantially better than for GJ 1132 or LHS\,1140, so we did not deem using [Ti/H] as a proxy for [Mg/H] or [Si/H] necessary for this target.)

\section{Radial velocity analysis} \label{sec:rv}

For {six of the nine} planets in our sample, we did not obtain any additional RVs for mass characterization. As such, for these targets, we take the RV semiamplitudes reported in the literature and recalculate our planet masses given our reanalyzed stellar masses calculated in Section \ref{subsec:stellchar-params}. We similarly recalculate the radii of all {nine} planets in our sample from $R_p/R_\star$ values in the literature and our reanalyzed stellar radii. In this section, we detail the process of modeling the RVs obtained for our three remaining targets selected for improved mass determination (GJ\,1132\,b, GJ\,1252\,b, and LTT\,3780\,b).

\subsection{Radial velocity model} \label{subset:rv-model}

To measure planetary parameters, we model all available RV time series for a given star simultaneously. The data are modeled as the sum of $N_p$ Keplerian planetary signals, where $N_p$ is the number of known planets in the system, and stellar variability. We model stellar  variability as a Gaussian process \citep[GP;][]{RasmussenWilliams2006} for each instrument separately (as our RV time series have been taken at a variety of wavelengths, and stellar variability is inherently chromatic), while sharing multiple hyperparameters between each GP. RV offsets $\gamma$ are added on a per-instrument basis as well.

We parameterized each Keplerian orbital signal with five parameters: the period of the planet $P$, the time of inferior conjunction $t_0$, the natural logarithm of RV semiamplitude $\log K$, and $h = \sqrt{e}\cos\omega$ and $k = \sqrt{e}\sin\omega$, where $e$ and $\omega$ are the orbital eccentricity and argument of periastron respectively \citep{LucySweeney1971}. All three planets of interest are expected to be tidally circularized given their short orbital period of $<2$ days, and so for these planets we restrict $h = k = 0$. Our three planets of interest all have extensive transit observations in the literature that provide exceptionally tight constraints on their linear ephemerides. Because the information content on $P$ and $T_0$ is dominated by these transit observations, we do not attempt to measure these parameters directly from the RV data and instead adopt informative priors on $P$ and $T_0$ from the transit analyses \citep{XueBean2024,KokoriTsiaras2023,BonfantiBrady2024}. We adopt uninformative, uniform priors on each planet's $\log{K}$. Our prior distributions are reported in Tables \ref{table:toi667-rv-table}, \ref{table:toi1078-rv-table}, and \ref{table:toi732-rv-table}.

Our GP models of stellar activity were computed using the \texttt{celerite2} package \citep{celerite1, celerite2}. We adopt the \texttt{SHOTerm} kernel, which is a common choice when modeling quasiperiodic RV variability due to active regions on a rotating star \citep[e.g.,][]{KrennKubyshkina2024,KroftBeatty2025}. The power spectral density of this kernel at a given frequency, $S(\omega)$, is given by

\begin{equation}
S(\omega) = \sqrt{\frac{2}{\pi}} \frac{S_0 \omega_0^4}{(\omega^2 - \omega_0^2)^2 + \omega_0^2\omega^2 / Q^2}
\end{equation}

for an undamped angular frequency $\omega_0$, a quality factor $Q$, and a scaling factor $S_0$. However, we use \texttt{celerite}'s reparameterizations of these hyperparameters, as they make the hyperparameters intuitive and map onto physical stellar parameters better. These reparameterized hyperparameters are the undamped period of the oscillator $\rho = 2\pi / \omega_0$, the damping timescale of the oscillator $\tau = 2Q/\omega_0$, and the standard deviation of the process (serving as a proxy for the GP amplitude) $\sigma = \sqrt{S_0 \omega_0 Q}.$ It should be noted that we fit $\log \tau$ and $\log \sigma$ instead of $\tau$ and $\sigma$ directly. As the oscillation period $\rho$ is directly informed by the rotation period of the star, we adopt priors for the undamped oscillation period of our GP equal to the previously reported rotation periods for each star (obtained from \citealp{BonfilsAlmenara2018}, \citealp{ShporerCollins2020}, and \citealp{CloutierEastman2020} for GJ\,1132, GJ\,1252, and LTT\,3780 respectively). $\log \tau$ was restricted by a uniform prior to be between $\log(2)$ and $\log(100)$ times the median of our prior for the rotation period of the star. $\log \sigma$ parameters were given uninformative uniform priors.

Each RV time series is modeled with its own GP; the hyperparameters $\rho$ and $\log \tau$ are shared between the individual GPs, but $\log \sigma$ is fit separately for each. GJ\,1132 and GJ\,1252\,both have three RV time series (the NIRPS RVs, the archival HARPS RVs, and the HARPS RVs taken simultaneously with NIRPS); due to the large temporal gap between the archival HARPS RVs and the HARPS measurements taken simultaneously with NIRPS, we treat them as two different datasets and instruments for this analysis for both stars. We fit six RV time series for LTT\,3780 (the NIRPS RVs, the CARMENES RVs, the archival HARPS RVs, the HARPS-N RVs, and the MAROON-X RVs, which are split up into the red and blue arms due to their chromatic nature.) All of these time series are described in more detail in Section \ref{sec:obs}.

We use the \texttt{emcee} package \citep{Foreman-MackeyHogg2013} to perform Markov chain Monte Carlo (MCMC) sampling of the joint posterior PDFs of our models. We use 64 walkers and run the MCMC for a sufficient number of steps until the chain length is more than $50\times$ the autocorrelation time, indicating the chains are well mixed.

\subsection{Results} \label{subsec:rv-results}

\begin{figure}
    \centering
    \includegraphics[width=0.95\linewidth]{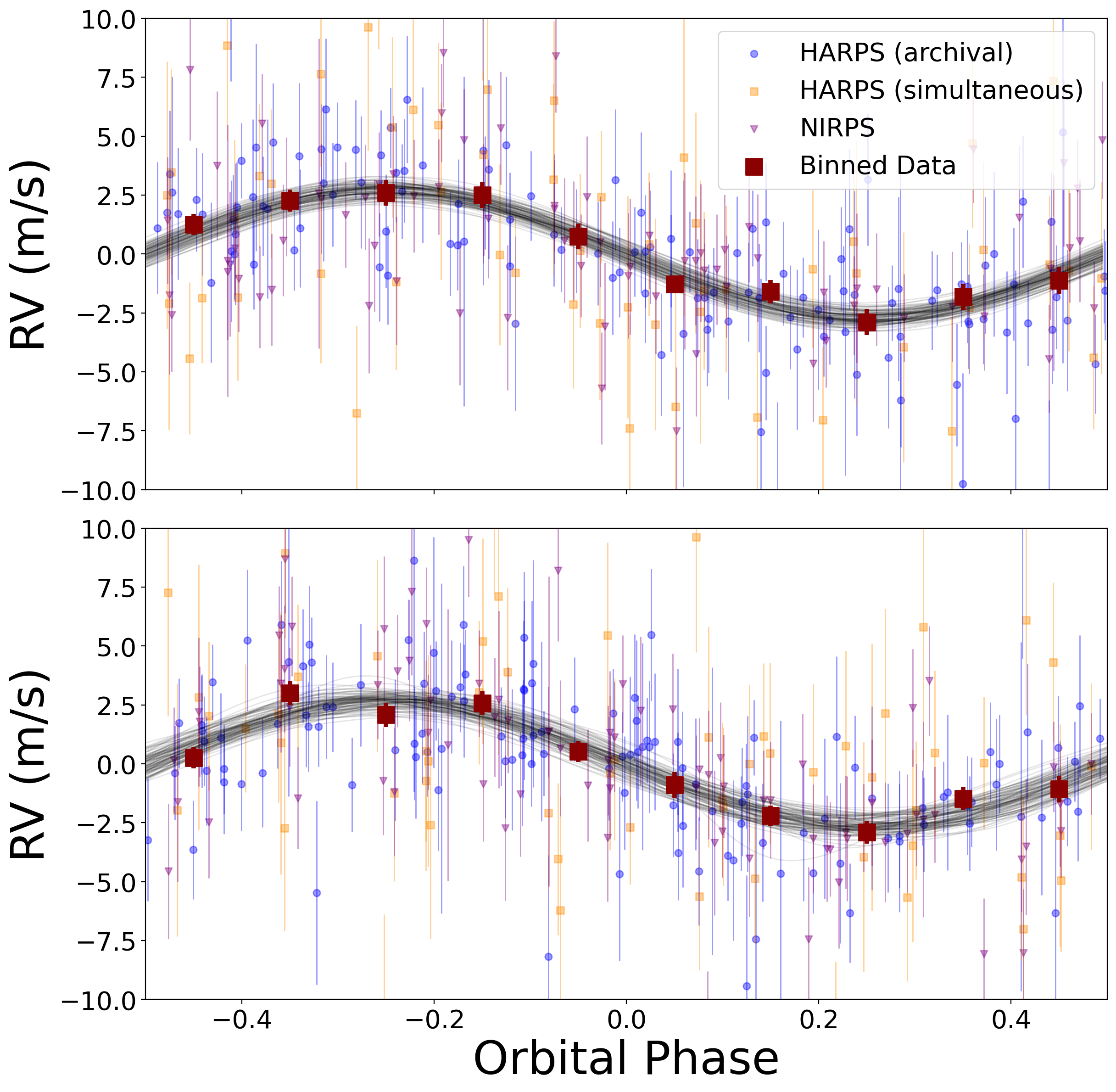}
    \caption{The detrended and phase-folded RV time series of GJ\,1132\,b (top) and GJ\,1132 c (bottom). Different draws of the Keplerian models from our posterior are superimposed in black. The different colored markers correspond to different RV time series. Binned data points are shown as red squares.}
    \label{fig:gj1132-phasefold}
\end{figure}

\begin{figure}[!ht]
    \centering
    \includegraphics[width=0.95\linewidth]{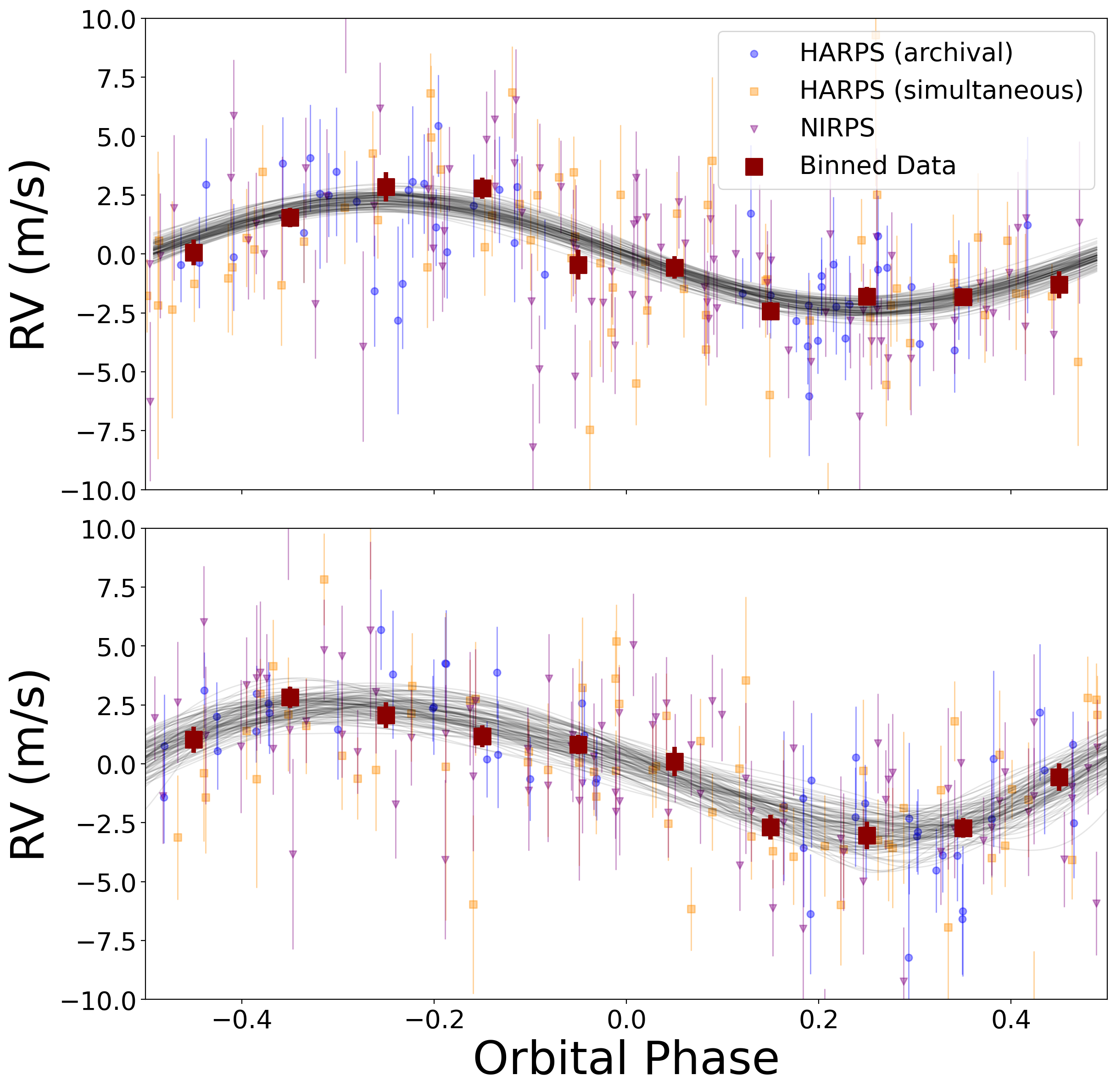}
    \caption{Phase-folded and detrended RV time series of GJ\,1252\,b (top) and GJ\,1252\,(c) (bottom). The best-fit Keplerian model is superimposed in black. The different colors and markers correspond to different RV time series. Binned data points are shown as red squares.}
    \label{fig:gj1252-phasefold}
\end{figure}

\begin{figure}
    \centering
    \includegraphics[width=0.95\linewidth]{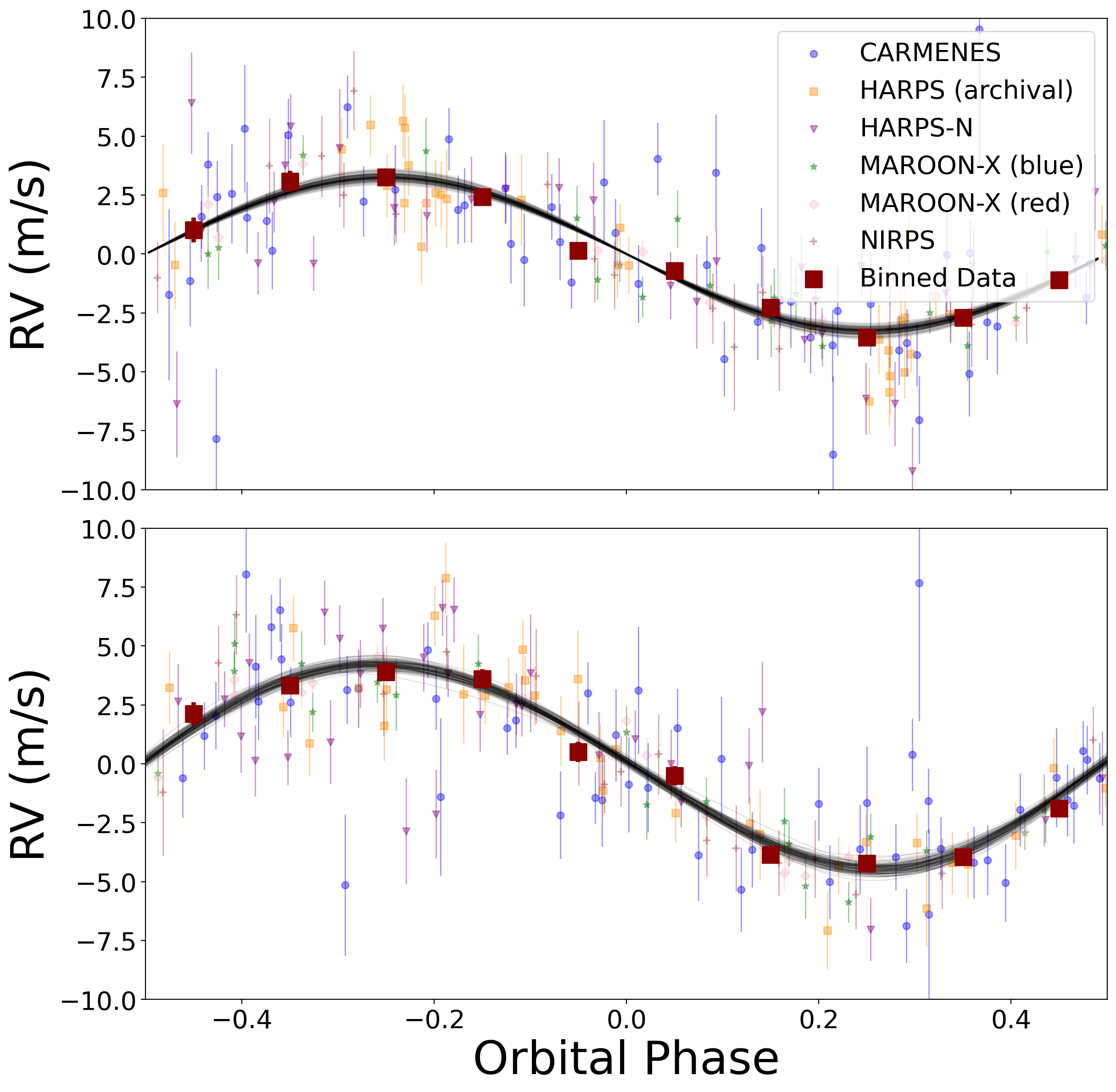}
    \caption{Phase-folded and detrended RV time series of LTT\,3780\,b (top) and LTT\,3780 c (bottom). The best-fit Keplerian model is superimposed in black. The different colors and markers correspond to different RV time series. Binned data points are shown as red squares.}
    \label{fig:ltt3780-phasefold}
\end{figure}

A table of the priors and posteriors for the RV model for each planetary system can be found in Tables \ref{table:toi667-rv-table}, \ref{table:toi1078-rv-table}, and \ref{table:toi732-rv-table}. Phase-folded plots of the planetary signals from each system are shown in Figures \ref{fig:gj1132-phasefold}, \ref{fig:gj1252-phasefold}, and \ref{fig:ltt3780-phasefold}; plots of the full RV time series for all three systems are shown in Section \ref{appendix:rv-plots}.

We obtain final masses of $1.69 \pm 0.15\,M_\oplus$ and $2.91 \pm 0.27 M_\oplus$ for GJ\,1132\,b and c, respectively. These are both consistent with the most recently reported masses of $1.66 \pm 0.19\,M_\oplus$ and $2.64 \pm 0.44\,M_\oplus$ for the two planets, identified by \citet{XueBean2024} and \citet{BonfilsAlmenara2018}, but represent significant mass precision improvements (from $9.7\sigma$ to $11.3\sigma$ for GJ\,1132\,b and $6\sigma$ to $10.8\sigma$ for GJ\,1132 c.)

We obtain final masses of $1.54 \pm 0.18 M_\oplus$ and $6.08 \pm 0.86 M_\oplus$ for GJ\,1252\,b and c, respectively. These are both consistent with the most recently reported masses of $1.32 \pm 0.28 M_\oplus$ and $7.4 \pm 1.0$ values for the two planets, identified in \citet{LuquePalle2022} (with GJ\,1252\,(c)'s previously estimated mass calculated from the reported semiamplitude and period), with GJ\,1252\,b's mass precision improving from $4.7\sigma$ to $8.6\sigma$, and GJ\,1252\,(c)'s mass precision remaining mostly unchanged (from $7.4\sigma$ to $7.1\sigma$, mostly due to a lower calculated mass.)

We obtain final masses of $2.34 \pm 0.10 M_\oplus$ and $7.89 \pm 0.26 M_\oplus$ for LTT\,3780\,b and c, respectively. These are both consistent with the most recently reported masses of $2.46 \pm 0.19 M_\oplus$ and $8.04_{-0.48}^{+0.50} M_\oplus$ for the two planets, reported in \citet{BonfantiBrady2024}, and represent mass precision improvements from $13\sigma$ to $23\sigma$ for LTT\,3780\,b and $17\sigma$ to $30\sigma$ for LTT\,3780\,c.

\subsection{The GJ\,1252 System and modeling GJ\,1252\,(c)} \label{subsec:rv-gj1252}

\citet{LuquePalle2022} reported the discovery of a second {nontransiting} planet candidate, {GJ\,1252\,(c)}, at 17.5 days. The inclusion of additional planets in an RV model can potentially impact the recovered masses of confirmed planets, even if the two planets do not have similar periods \citep[e.g.,][]{AlmenaraBonfils2022}. Here we consider RV models of the GJ\,1252 system with the inclusion of  one versus two planetary signals, and perform a detailed model comparison between the two.

\begin{figure}
    \centering
    \includegraphics[width=0.95\linewidth]{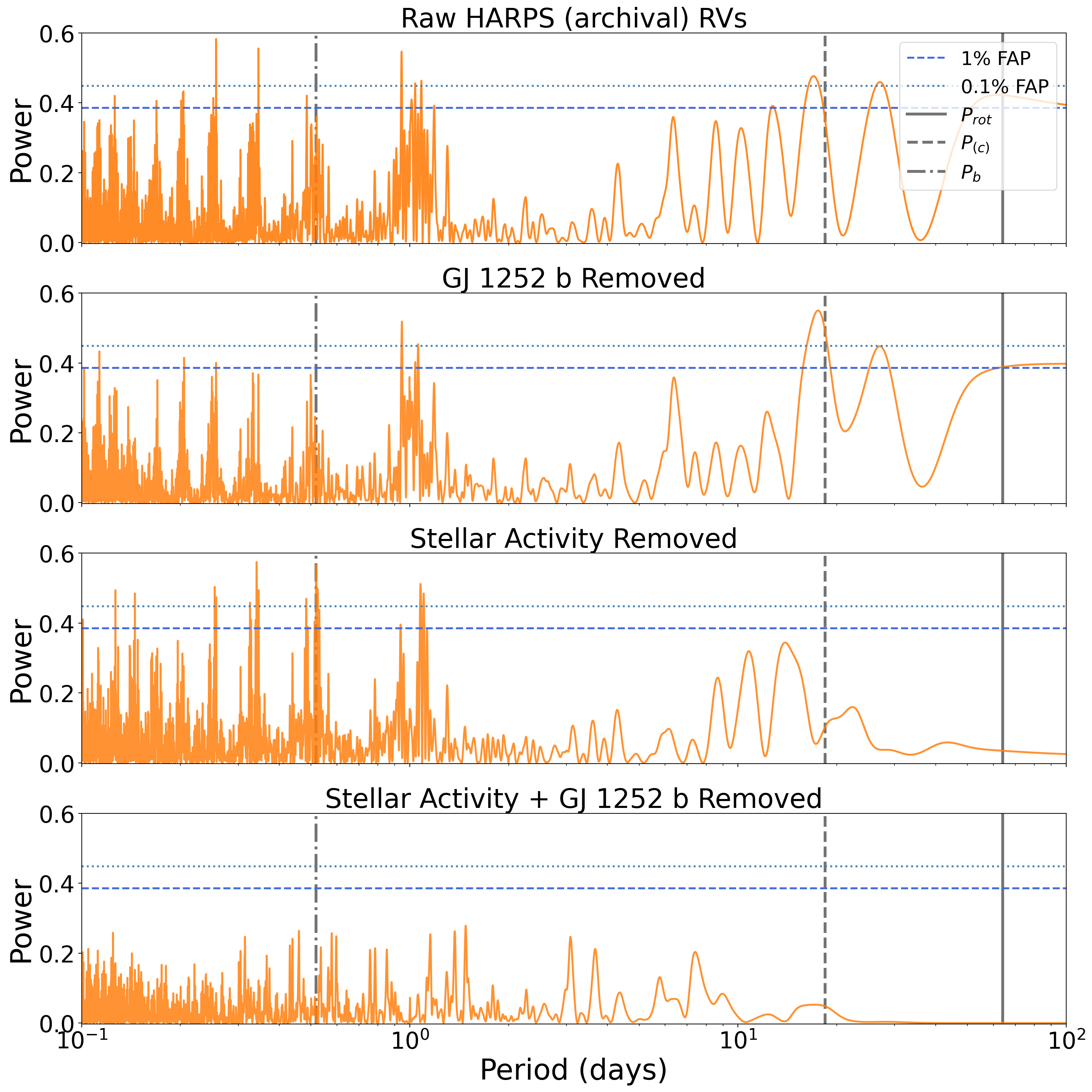}
    \caption{Generalized Lomb-Scargle periodogram of the archival HARPS RVs of GJ\,1252. Each panel shows a different step of the 1-planet model: the raw RVs (top); the RVs with the maximum a-posteriori (MAP) GJ\,1252\,b Keplerian model subtracted out (second); the RVs with the MAP stellar activity model subtracted out (third), and the RVs with the MAP stellar activity model and GJ\,1252\,b Keplerian model both subtracted out (bottom). Red lines corresponding to false alarm probabilities (FAPs) of $1\%$ and $0.1\%$ are overplotted, as are gray lines corresponding to $0.52$d (GJ\,1252\,b's period), $17.5$d (GJ\,1252\,(c)'s period reported in \citet{LuquePalle2022}), and $64$\,d (the stellar rotation period reported in \citet{ShporerCollins2020}).}
    \label{fig:gj1252-gls}
\end{figure}

We model both the 1-planet case (with only GJ\,1252\,b on a circular orbit) and the 2-planet case, which includes the planet candidate GJ\,1252\,(c) on a freely eccentric orbit. Both of these models include informed stellar activity GPs with the same priors. We report the results of both GJ\,1252 model fits in Table \ref{table:toi1078-rv-table}. We note that the data is effectively modeled in both cases. While the 2-planet case models the data well, we see no significant $17.5$d signal in the residuals of the archival HARPS RVs, as can be seen in Figure~\ref{fig:gj1252-gls}. We additionally see no $17.5$d signal in the NIRPS RVs, either the raw RVs or the residuals. In addition, the rotation period that both cases independently fit to is $55$d, at the $3:1$ harmonic of the $18.41$d period of GJ\,1252\,(c) identified in the posterior, strongly suggesting that either this signal is not planetary in nature or that the stellar activity GP can effectively model out this signal.

To identify the preferred model, we estimate the Bayesian evidence of our models using the Perrakis estimator \citep{Perrakis2014}. The Perrakis estimator is an estimator of the fully marginalized likelihood (a.k.a the Bayesian evidence) that utilizes the unnormalized posterior PDF from our MCMC sampling as an importance sampler. Assuming that our MCMC sampling has adequately explored the posterior parameter space for each model, the Perrakis estimator can approximate the Bayesian evidence ratio of competing models with a precision of roughly $10^3$ \citep{Nelson2020}, meaning that Bayesian evidence ratios produced by the Perrakis estimator that exceed $\sim10^3$ are generally considered significant. This method has been commonly used in recent years to estimate Bayesian evidences of RV planetary signals for the purpose of model comparison \citep[e.g.,][]{DiazSegransan2016,BonomoDesidera2017,Lillo-BoxFigueira2020,SuarezMascarenoFaria2020,CloutierGreklek2023,Balsalobre-RuzaLillo-Box2025}.

We find that the 2-planet model is favored over the 1-planet model by $\Delta\ln(Z) = 26.6$, which would appear to provide very strong evidence for the 2-planet model. However, we reemphasize that our 1-planet+GP model does effectively fit the 17.5 day periodicity seen in the GLS of the GJ\,1252 RVs such that we do not claim to have evidence that the 17.5-day signal is planetary in nature. We also note that the resulting masses for GJ\,1252\,b from the two models are consistent (i.e., $1.43\pm 0.19 M_\oplus$ and $1.54 \pm 0.18 M_\oplus$ for the 1- and 2-planet models, respectively). Consequently, the results from our forthcoming interior structure analysis will not be significantly impacted by our choice of model. We elect to use the GJ\,1252\,b mass obtained from our 2-planet model in subsequent analyses given that it is strongly favored by the Bayesian evidence ratio. 

\subsection{Mass-radius diagram}\label{subsec:rv-mr}

\begin{figure}[h]
    \centering
    \includegraphics[width=0.95\linewidth]{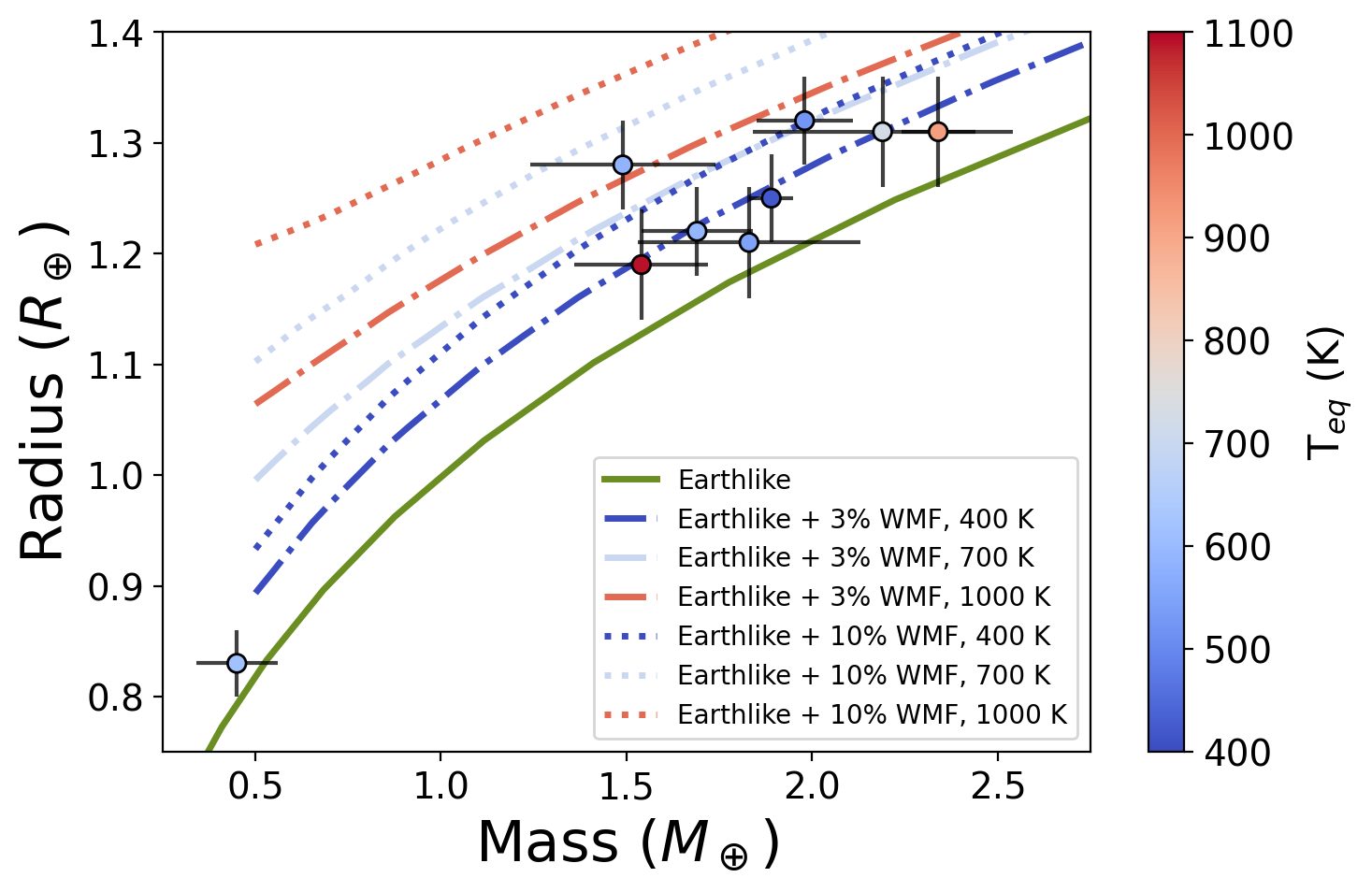}
    \caption{A plot detailing the reanalyzed radii vs. masses of all the planets in our sample. The colors of data points are determined based on {equilibrium temperature}, according to the colorbar. Several composition curves are overlaid. A curve corresponding to an Earth-like composition from \citet{ZengJacobsen2019} is shown as solid green line. Curves corresponding to water mass fractions of $3\%$ (dash-dotted) and $10\%$ (dotted) are shown at 400, 700, and 1000 K, and are linearly interpolated based on the same pre-tabulated models discussed in Section \ref{subsec:disc-wmf}.}
    \label{fig:mr}
\end{figure}

Table \ref{table:planet-params} details the reanalyzed masses and radii of our entire sample of planets. We have plotted the masses and radii of our planets, {with} an Earth-like composition curve from \citet{ZengJacobsen2019} and composition curves taking into account water dissolution and sequestration (discussed in Section \ref{subsec:disc-wmf}), in Figure \ref{fig:mr}.

{All nine planets lie above the Earth-like composition curve in mass-radius space, indicating radii larger than expected for Earth-like compositions at their measured masses, and therefore lower bulk densities. We also note that none of our planets lie below this curve, meaning we do not see any super-Mercuries in our sample.}

\section{Core mass fraction determination} \label{sec:cmf}

We derive the CMFs of our planets in two different ways for the purpose of comparison. We begin by calculating each planet's CMF from its measured mass and radius, and assuming that its bulk composition is well-described by just two components: a differentiated iron core beneath a silicate mantle (CMF$_{planet}$). Secondly, we calculate each planet's expected CMF based on the Mg, Si, and Fe abundances of its host star (CMF$_{star}$). In calculating CMF$_{star}$, we are assuming that the planets' refractory abundance ratios are identical to its host star's. 

Comparing the CMF$_{planet}$ and CMF$_{star}$ values for each planet may give us insight into the formation and evolutionary processes of close-in super-Earths around M dwarfs. {Specifically, if the planetary and stellar CMF predictions are roughly equal at the population level, then that would imply that the compositions of terrestrial super-Earths around M dwarfs are set by the stellar composition and that no subsequent iron enhancement or depletion processes are affecting these planets' compositions. If these planets instead tend to exhibit larger CMFs than expected from their host stars, then that would imply a more massive iron core than is expected from formation, and suggests that mantle stripping by giant impacts is a driving process that sets these planets' compositions \citep{MarcusSasselov2010}. Lastly, if these planets tend to exhibit smaller CMFs than expected from their host stars, that would imply either an iron depleted interior or some form of dilution by a low density species (e.g., water). Planet formation models do not predict significant iron depletion of planets \citep[e.g.,][]{ScoraValencia2020}, as the refractory abundances in FGK stars in the solar neighborhood are largely homogeneous \citep{BrewerFischer2016,BedellBean2018}. In addition, the relevant species found inside terrestrial planets have very similar condensation temperatures \citep{DornHarrison2019}, and so they should contribute in nearly equal proportions to forming protoplanets. Because of this, if the planets in our sample have smaller CMFs than expected, the assumption of a planet composed entirely of iron and silicates is likely incorrect. Recent theory suggests that super-Earths can sequester water deep inside their cores, inflating their radii, reducing their bulk densities and inferred CMF values compared to when a pure iron core is incorrectly assumed \citep{LuoDorn2024}. However, for planets to have significant amounts of water sequestered inside their cores, the water would have to be accreted early on in the planet formation process. Smaller CMF values than expected would therefore be consistent with water-rich formation around M dwarfs and help constrain the nature of the accretion process.}

\begin{figure*}
    \centering
    \includegraphics[width=\linewidth]{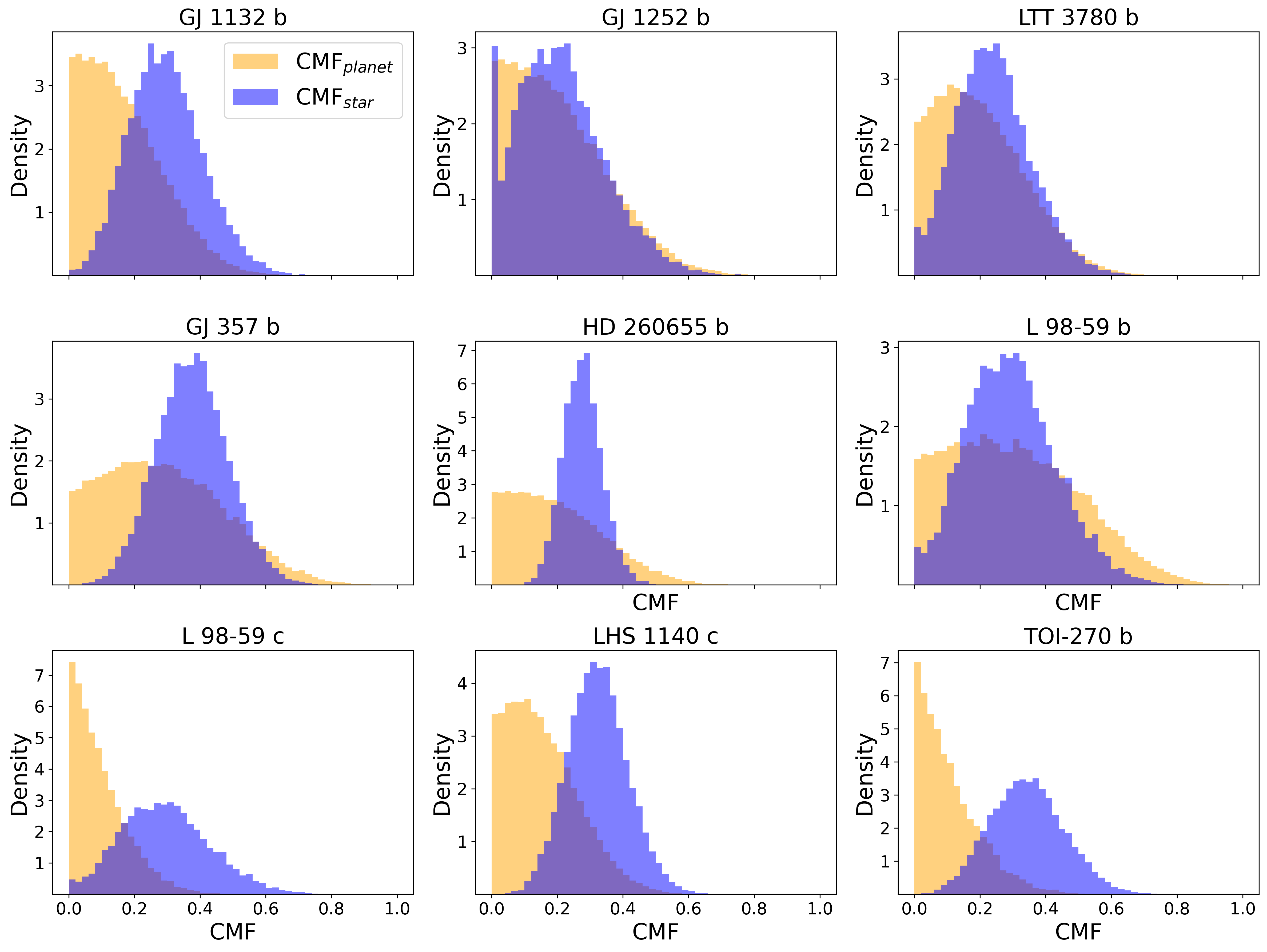}
    \caption{A set of posterior PDFs showing the distributions of CMF$_{planet}$, the CMF of the planets as measured from their masses and radii (orange), and CMF$_{star}$, the equivalent CMF as inferred from the refractory abundances of their host stars (blue), for each planet. Due to the high uncertainty in abundances for some stars, such as GJ\,1252, some draws from the posterior are fit with very low [Fe/H] values and very high [Mg/H] and [Si/H] values, resulting in a very low CMF$_{star}$, causing a large amount of posterior draws to converge to CMF$_{star}$ $\approx 0$.}
    \label{fig:CMF-hists}
\end{figure*}

\subsection{Mass and radius-derived core mass fraction estimates} \label{subsec:cmf-mr}

We use the rocky planet interior structure modeling package \texttt{exopie} \citep{PlotnykovValencia2024} to infer CMF$_{planet}$ values from the planetary masses and radii reported in Table \ref{table:planet-params}. \texttt{exopie} is an updated version of the earlier internal structure code \texttt{SuperEarth} \citep{PlotnykovValencia2020}. We use \texttt{exopie} to generate forward models of exoplanet radii given an exoplanet mass, CMF, as well as the molar fraction of silicon in the core and the molar fraction of iron in the mantle, \texttt{xSi} and \texttt{xFe}, which are allowed to vary uniformly between $0$ and $0.2$. As these planets are very close to their host stars, with $T_\textrm{eq} > 373\,\textrm { K}$ (the boiling point of water at 1 atm), our models assume that these planets have no water or significant atmospheres, meaning these models assume interior compositions that include a primarily iron core and a primarily silicate mantle. {These assumptions are appropriate for likely rocky planets below the radius valley \citep[$R < 1.6 R_\oplus$,][]{Rogers2015}; as all of our planets have a radius below $1.4 R_\oplus$, this is a safe assumption for our entire sample.}

\texttt{exopie} samples the joint prior parameter space on ${M_p, \texttt{xSi}, \texttt{xFe}, \texttt{CMF}}$ to find the {set of planetary parameters that result in a planet whose radius} best fits the observed planet radius{, assuming realistic mineralogy}. The corresponding CMF posterior is our parameter of interest (i.e., CMF$_{planet}$). For each planet, we sample the parameter space with $10^5$ draws to quantify the CMF$_{planet}$ posterior. The resulting CMF$_{planet}$ distributions are depicted in Figure~\ref{fig:CMF-hists} and point estimates are reported in Table~\ref{table:planet-params}.

\subsection{Stellar abundance-derived core mass fraction estimates} \label{subsec:cmf-stellar}

In addition to inferring CMF$_{planet}$ from the planetary masses and radii, here we compute the planetary CMFs that are expected from the refractory abundances of their host stars (i.e., CMF$_{star}$). The CMF$_{star}$ values assume that rocky planets preserve the refractory abundance ratios exhibited in the host stellar photosphere such that discrepancies between CMF$_{planet}$ and CMF$_{star}$ can tell us something about the rocky planet formation process.

{We use \texttt{exopie} to derive CMF$_{star}$ from stellar refractory abundances, using an equivalent chemical planetary model to that used in the interior structure modeling presented in Section \ref{subsec:cmf-mr}. \texttt{exopie} solves for the chemical inventory of a sampled rocky configuration, allowing \texttt{xSi} and \texttt{xFe} to vary between 0 and 0.2 as in Section \ref{subsec:cmf-mr}, such that the final refractory ratios (Fe/Mg, Fe/Si, Mg/Si) of the planet best match the measured stellar abundance ratios in Table \ref{table:stellar-params}. This calculation is very similar to previous estimates to calculate a CMF from stellar abundances \citep[e.g.,][]{SchulzeWang2021}; in fact, when \texttt{xSi} and \texttt{xFe} are set to 0, the equivalent expression for CMF$_{star}$ is functionally equivalent to that presented in Equation (2) of \citet{SchulzeWang2021}, with any differences in estimates due only to the different mineralogy used in \texttt{exopie} versus what is used in that work.}

For each star, we draw $10^4$ samples of the refractory abundances of Mg, Si, and Fe from assumed Gaussian distributions set by the mean values and uncertainties identified in Section \ref{subsec:stellchar-abund} {and calculate the corresponding CMF$_{star}$ with \texttt{exopie}}. The resulting CMF$_{star}$ distributions are included in Figure~\ref{fig:CMF-hists}, with point estimates also reported in Table~\ref{table:planet-params}.

\section{Discussion} \label{sec:disc}

\subsection{The CMF$_{planet}$ and CMF$_{star}$ distributions are statistically different} \label{subsec:disc-statistics}

We find that for {eight of our nine} planets in our sample (excepting only L\,98-59\,b), the median values of the posteriors for CMF$_{planet}$ is lower than the median values for CMF$_{star}$. While the large uncertainties in both CMF distributions make it difficult to draw concrete conclusions for any individual system, our finding that CMF$_{planet}$ < CMF$_{star}$ applies to the majority of planets in our sample suggests a clear trend.

\begin{figure}
    \centering
\includegraphics[width=0.95\linewidth]{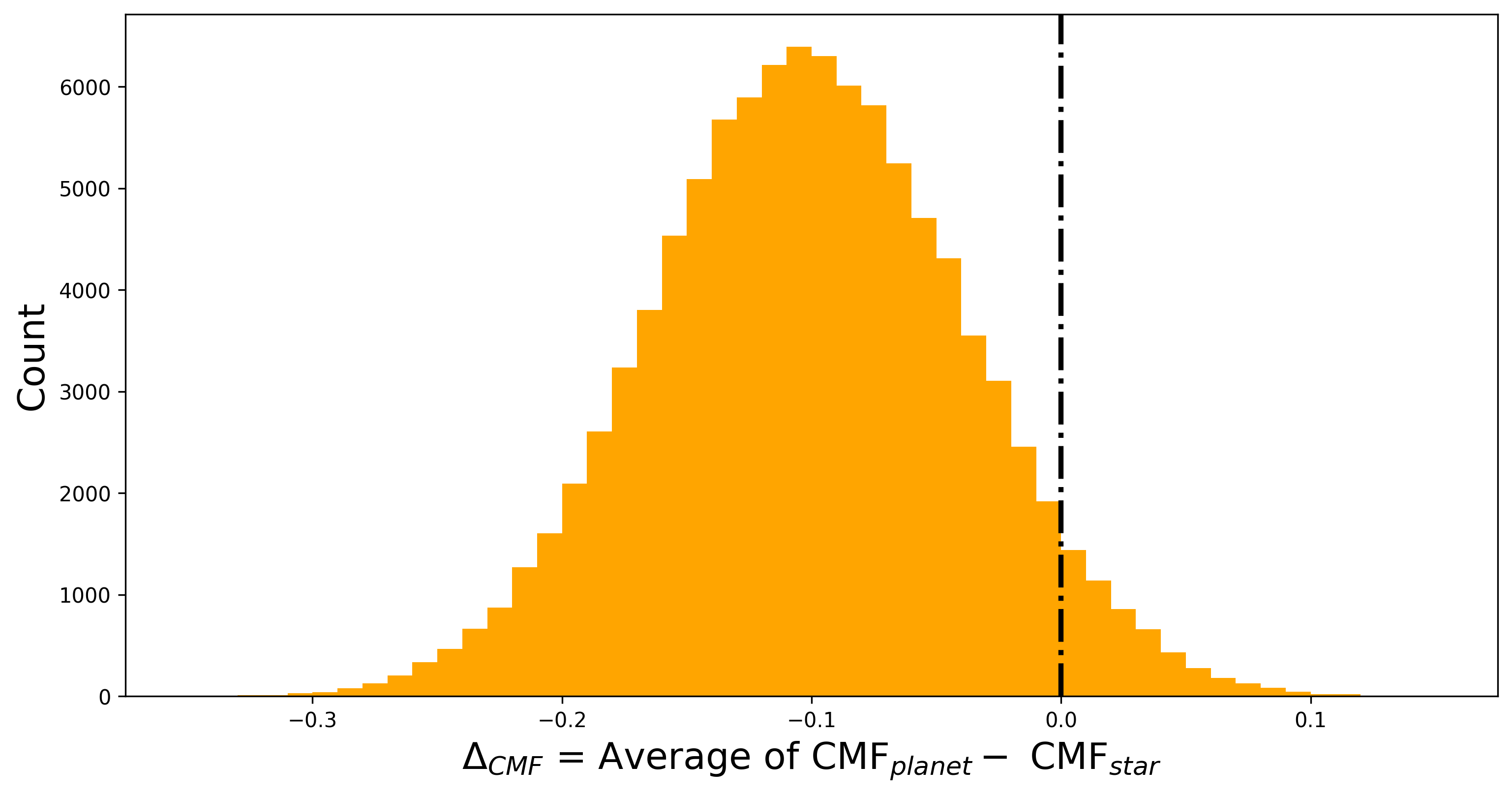}
    \caption{{The distribution of $\Delta_{CMF}$ draws, as described in Section \ref{subsec:disc-statistics}. A vertical black line corresponding to $\Delta_{CMF} = 0$ is overplotted.}}
    \label{fig:delta-CMFs}
\end{figure}

{We quantify this by estimating population mean difference between our two CMF estimates, $\Delta_{CMF}=$ CMF$_{planet}$ $-$ CMF$_{star}$, through Monte Carlo resampling. To estimate $\Delta_{CMF}$, we sample draws from the posteriors of CMF$_{planet}$ and CMF$_{star}$ for each of our planets. We then calculate the difference between the mean of the CMF$_{planet}$ draws and the CMF$_{star}$ draws (which is equivalent to calculating the average of the differences between the two CMF estimates for each planet). We repeat this process $10^5$ times and observe the distribution of $\Delta_{CMF}$ draws, which are shown in Figure \ref{fig:delta-CMFs}}.

{We find that $94.7\%$ of $\Delta_{CMF}$ draws are below 0, while $5.3\%$ of draws are above 0. $95\%$ of $\Delta_{CMF}$ draws fall between $[-0.224, 0.022]$. Because the vast majority of the $\Delta_{CMF}$ distribution lies below 0, we interpret this as strong evidence of a systematic offset between our CMF$_{star}$ and CMF$_{planet}$ distributions at the population level.}

\subsection{possible interpretations of planetary structure} \label{subsec:disc-interp}

We have established that the CMFs of hot super-Earths around M dwarfs in our sample are systematically lower than expected if they formed solely from the primordial refractory materials in their host stars. If we continue to assume that these planets are constructed out of solely iron and silicates, then our results would imply that they have significantly less massive iron cores than their host stars' abundances would suggest. Planet formation models do not predict significant iron depletion of super-Earths \citep[e.g.,][]{ScoraValencia2020}. This follows from the observational result that the refractory abundances of FGK stars in the solar neighborhood are largely homogeneous \citep{BrewerFischer2016,BedellBean2018} and because Mg, Si, and Fe all have similar condensation temperatures \citep{DornHarrison2019}. As these stars aren't expected to be iron-depleted, and there is a vanishingly small temperature range in a protoplanetary disk where some but not all of these three refractories would have condensed, we do not expect these planets to have formed from iron-depleted material. 

If they exist, thick atmospheres on these planets (either hydrogen/helium atmospheres or steam atmospheres) could in theory increase the observed radii of these planets, decreasing their measured densities and thus accounting for this discrepancy. However, as these planets are very hot and relatively small, thick atmospheres do not appear likely on these planets. Hydrogen and helium would be likely to have undergone atmospheric escape, while if a differentiated water atmosphere were to have existed on these planets, their high equilibrium temperatures would have likely led any steam to photodissociate, whereupon the H$_2$ and O$_2$ would likely be lost to space \citep{SchaeferWordsworth2016}. It should be noted as a caveat that some O$_2$ may be sequestered in this instance; see \citealp{SchaeferWordsworth2016} for a discussion on this focusing on GJ\,1132\,b specifically. It is also possible that water on these planets may have been converted to hydrogen or sequestered as well, resulting in low atmospheric water mass fractions (WMFs) \citep{LuoDorn2024,GuptaLuo2025,WerlenDorn2025}.

Models based on observations of XUV flux from M dwarfs predict that most super-Earths around M dwarfs (including specifically our entire sample of planets) are unlikely to retain significant atmospheres, as well \citep{PassCharbonneau2025}. Observational evidence supports this too: observations of super-Earths around M dwarfs are unlikely to find significant atmospheres \citep{KreidbergStevenson2025}. In addition, observations attempting to directly constrain the atmospheres of several planets in our sample have been performed, and have found results consistent with no thick atmosphere for GJ\,1132\,b \citep{LibbyRobertsBertaThompson2022,MayMacDonald2023,BennettMacDonald2025,PalleYan2025}, {GJ\,1252\,b \citep{CrossfieldMalik2022},} LTT\,3780\,b \citep{AllenEspinoza2025}, GJ\,357\,b \citep{AdamsRedaiWogan2025,TaylorRadica2025}, {L\,98-59\,c \citep{ScarsdaleWogan2024},} and LHS\,1140\,c \citep{CadieuxDoyon2024,RochonArtigau2025}. L 98-59\,b is believed to potentially host a thin secondary atmosphere \citep{BelloArufeDamiano2025}, and TOI-270 b may host a thin steam atmosphere \citep{CoulombeBenneke2025}, but both are likely not thick enough to substantially affect the observed radius of this planet. As such, it seems unlikely that the presence of thick atmospheres is the primary cause of the difference between our CMF$_{planet}$ and CMF$_{star}$ distributions.

One alternative explanation is that the interiors of hot super-Earths around M dwarfs are made, in part, of a nonnegligible mass fraction of lower density materials, such as water. The structure models used in Section \ref{sec:cmf} to calculate the CMFs of our planets do not account for the presence of volatiles, either sequestered or differentiated, and so the presence of volatiles on the planets but not in our models could explain the discrepancy. Geochemical theory suggests that water can be sequestered deep inside the cores and mantles of super-Earths \citep{LuoDorn2024}. Water sequestration can substantially inflate the planetary radius by up to 25\% more than it would have without water, in the process deflating the bulk density and thus the inferred CMF$_{planet}$ \citep{LuoDorn2024}. Our results suggest that hot super-Earths around M dwarfs do harbor nonnegligible WMFs.

\subsection{A dearth of super-Mercuries} \label{subsec:disc-no-super-merc}

{Super-Mercuries are a proposed class of exoplanet characterized by radii similar to those of super-Earths but with very high densities, requiring CMFs well above that of Earth to explain.} We note that none of our targets exhibit a CMF$_{planet}$ value that exceeds CMF$_{star}$. We therefore observe no evidence of iron enhancement in any of our targets, which is the defining characteristic of super-Mercuries. An elevated CMF$_{planet}$ is an expected outcome of mantle stripping by giant impacts \citep{MarcusSasselov2010}, which is believed to have played a key role in the late phases of final assembly of the terrestrial planets in the Solar System \citep{Levison2003,Raymond2005,Kokubo2006,planetformation}. Our results suggest that giant collisions do not play a key role in the formation of super-Earths around M dwarfs, perhaps owing to their low occurrence rates of cold gas giants \citep{Fulton2021,BryantBayliss2023,Pass2023} that may be needed to perturb planetesimals onto crossing orbits \citep{Chambers1998}.

\subsection{Calculating the WMFs of hot super-Earths around M dwarfs} \label{subsec:disc-wmf}

It should be noted that while here we calculate WMFs assuming water-rich planets, alternative proposed compositions such as soot planets or other light elements in the core \citep[e.g.,][]{LiBergin2026} may be consistent with the planetary masses and radii reported in this work. However, we reemphasize that we favor the water-rich interpretation of given the wealth of observational evidence for water-rich formation already discussed in Section \ref{sec:intro}. With the knowledge that the presence of sequestered volatiles inside these planets is possible, then, here we aim to calculate the planets' WMFs assuming that the water is either sequestered in the planet's core and mantle or fully differentiated. 

We obtain our first set of WMF estimates by allowing the water to be sequestered into the planet's core and mantle. The partitioning of water depends on the planetary parameters mass, temperature, and compositional mass fractions, and are tabulated using the model presented in \citet{LuoDorn2024}. The models define the radius of the planet as the radius of 1 mbar pressure in any steam atmosphere, if it exists. For simplicity, we elect to fix the CMF to an Earth-like value of $0.325$ while sampling the mass, $T_\textrm{eq}$, WMF, and (as a dependent variable) radius over the ranges of $[0.5, 30] M_\oplus$, $[300, 2500]$\,K, $[0.01, 0.40]$, and roughly $[0.85, 3] R_\oplus$, respectively. We note that the CMF$_{star}$ values for all of our target stars are consistent with an Earth-like CMF to $<1\sigma$, such that fixing the CMF to CMF$_{\oplus}$ is a reasonable assumption. 

We {calculate this first set of WMF estimates by interpolating} our interior structure model table with $10^4$ samples from the input parameter posteriors. For each of the planets in our sample, we resample the mass, radius, and T$_\textrm{eq}$ from our median values and uncertainties in Table \ref{table:planet-params}, assuming Gaussian errors. We then perform a piecewise linear interpolation over those three parameters to find the WMF of this realization. All sampled parameters outside of the ranges of planet parameters sampled in the model tabulations were ignored. The resulting WMF posteriors are given in Table \ref{table:WMFs}.

We obtain our second set of WMF estimates by assuming that the planet is fully differentiated with the water layer forming the surface of the planet. While the high equilibrium temperatures of these planets may preclude a surface water layer, we still wish to measure the expected WMFs of these planets with this method as a point of comparison. {To calculate this second set of WMF estimates,} we use the interior structure modeling code \texttt{smint} \citep{PiauletBenneke2021,PiauletBenneke2024}, which couples an interpolator of planetary structure models of rocky \citep{ZengSasselov2016} and irradiated ocean worlds \citep{AguichineMousis2021} with the MCMC sampler \texttt{emcee} \citep{Foreman-MackeyHogg2013}. These planets are modeled as an iron core, a silicate rock mantle, and a water layer on top. The water layer is expected to be in a vapor/supercritical phase, as all our targets have $T_\textrm{eq} > 400 \textrm{ K}$ \citep{AguichineMousis2021}. The planetary radius produced by the models for a given mass and WMF is compared to the observed radius; the mass and irradiation temperatures $T_\textrm{irr}$ are constrained by Gaussian priors from the values found in Tables \ref{table:planet-params}. The $T_\textrm{irr}$ values are assumed to be equal to the $T_\textrm{eq}$ values, which are calculated as if the Bond albedo is 0. To be able to properly compare to our first set of WMF estimates, we fix the CMF to the same value of $0.325$. The resulting WMFs are reported in Table \ref{table:WMFs}.

We find that for nearly every planet in our sample, the resulting WMFs assuming water is sequestered are consistent with $0.01$ (the smallest WMF sampled in the pre-tabulated models.) The only planet for which this is not the case is L\,98-59\,b, though it should be noted that due to its small mass, $70\%$ of the sampled parameters were outside of the range of input parameters for the pre-tabulated models, and so were ignored when calculating the posterior. Similarly, we find that the WMFs for the planets in our sample tend to hover around $1-2\%$ if water is differentiated on the surface of the planet. These numbers, while largely not consistent with WMFs of $0$, are nevertheless very low.

\begin{table}[]
\caption{Calculated WMFs}
\begin{tabular}{c|c|c}
\hline \hline
Planet & WMF (sequestered) & WMF (differentiated) \\ \hline
GJ\,1132\,b & $<0.07$ & $0.02 \pm 0.01$\\
GJ\,1252\,b & $<0.03$ & $0.012_{-0.007}^{+0.009}$\\
LTT\,3780\,b & $<0.04$ & $0.013_{-0.008}^{+0.010}$ \\
GJ\,357\,b & $< 0.08$ & $<0.04$ \\
HD\,260655\,b & $<0.11$ & $0.02_{-0.01}^{+0.02}$ \\
L\,98-59\,b & $0.023 \pm 0.004$ & $0.005 \pm 0.003$\\
{L\,98-59\,c} & {$<0.14$} & {$0.03 \pm 0.01$} \\
LHS\,1140\,c & $<0.12$ & $0.02 \pm 0.01$\\
TOI-270\,b & $<0.19$ & $0.04 \pm 0.01$ \\ \hline \hline
\end{tabular} \label{table:WMFs}
\smallskip
    \footnotesize

    The distributions of both sets of WMF estimates. 95th percentile upper limits on the WMF estimates are reported instead of error bars where the WMF posterior is consistent with $0$ for the differentiated WMFs, or $0.01$ for the sequestered WMFs.
\end{table}

Recently, \citet{RogersDorn2025} compared the interior models of \citet{LuoDorn2024} with an observed sample of super-Earths to place upper limits on the WMFs of super-Earths at the population level. In the case of super-Earths without atmospheres or mantle outgassing, they find that super-Earths have no more than 3\% WMFs on average.

We find that the WMFs of our sample planets calculated using both methods are largely consistent with this WMF upper limit. We may not necessarily expect water on these planets to be differentiated, and photodissociation or sequestration of this water seems more likely instead \citep{WerlenDorn2025}; however, measuring the WMFs of these planets alone does not seem sufficient to distinguish between these two possibilities.

\subsection{The star-planet compositional connection} \label{subsec:disc-otherstudies}

Past studies have tried to quantify the relationship between the compositions of rocky planets and their host stars \citep{PlotnykovValencia2020,AdibekyanDorn2021,SchulzeWang2021,LiuNi2023,AdibekyanDeal2024,BrinkmanPolanski2024,BrinkmanWeiss2025,PlotnykovValencia2026}. These studies have produced a range of different results, in part due to differences in sample selection and methodologies. For example, \citet{PlotnykovValencia2020} found that some planets may be iron-depleted relative to their host stars; \citet{BrinkmanPolanski2024} identified {a relationship consistent with a one-to-one relationship between planetary and host CMFs, though due to uncertainties in both was unable to draw a conclusive correlation}, while \citet{AdibekyanDorn2021} identified a very strong correlation between the planetary and host stellar iron mass fractions. However, all of these studies were primarily based on analysis of planetary systems around FGK stars, with little to no contributions from M dwarfs. In this study, however, we are in a position to compare the star-planet correlations reported for FGK stars to that around M dwarfs.

\begin{figure}
    \centering
    \includegraphics[width=0.95\linewidth]{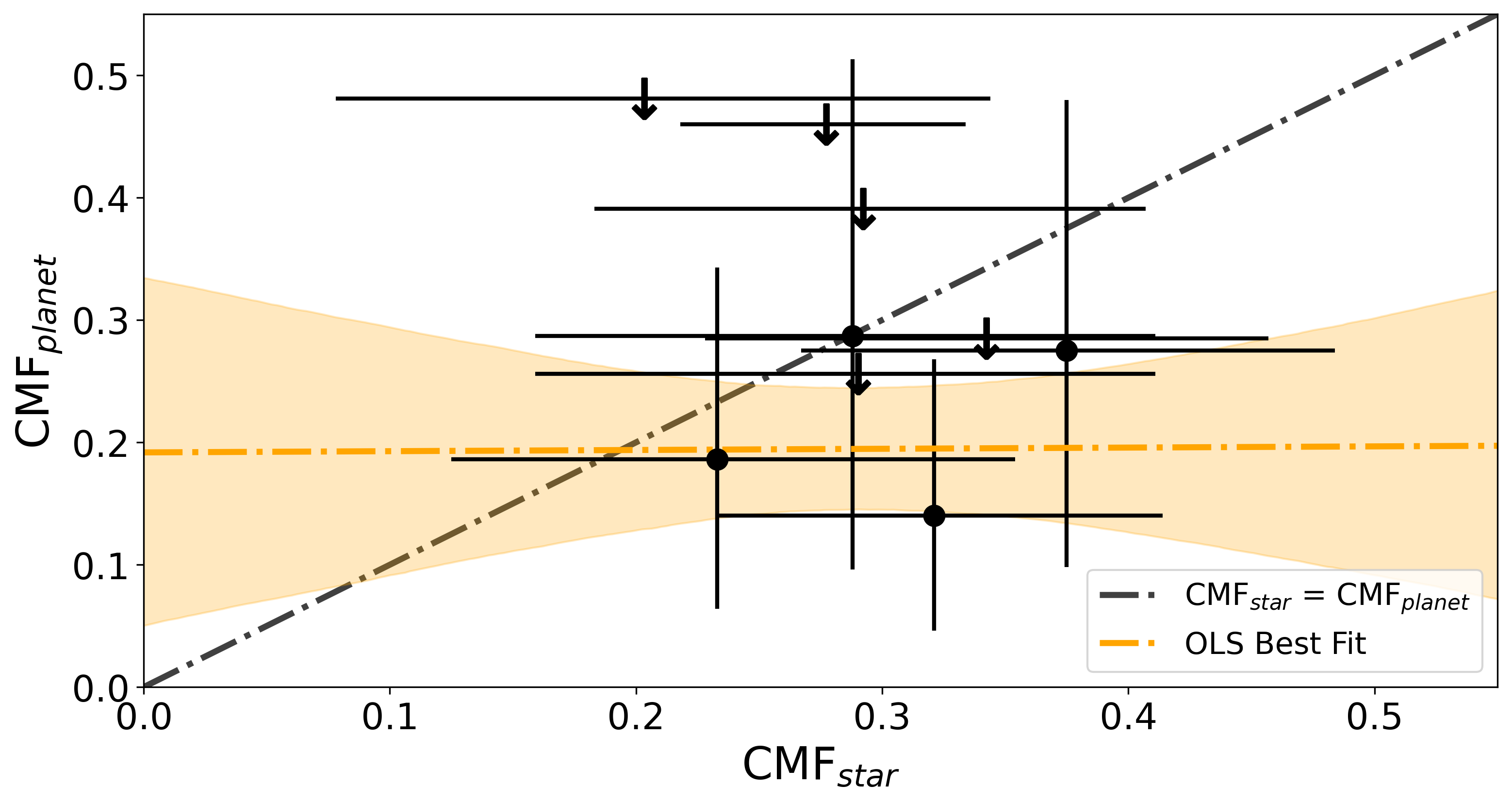}
    \caption{The correlation between CMF$_{planet}$ and CMF$_{star}$ for the {nine} hot super-Earths in our sample. $95\%$ upper limits are denoted with downwards arrows. The line CMF$_{planet}$ $=$ CMF$_{star}$ is overplotted as a black dash-dotted line. Our OLS fit with $1\sigma$ confidence intervals is shown in orange. The ODR fit is not shown here due to its high uncertainty.}
    \label{fig:CMF-scatter}
\end{figure}

Several of these studies \citep{AdibekyanDorn2021,AdibekyanDeal2024,BrinkmanPolanski2024} fit a linear relationship between their two CMF estimates (or proxies thereof), which we replicate here for our M dwarf planet sample. To quantify the relationship between CMF$_{star}$ and CMF$_{planet}$, we perform a linear fit of the form $\textrm{CMF}_{planet} = m\textrm{CMF}_{star}+b$. We perform a Monte Carlo simulation by directly resampling CMF values from the distributions reported in Table~\ref{table:planet-params} and shown in Figure~\ref{fig:CMF-hists} $10^4$ times and perform an ordinary least-squares (OLS) fit for each realization. We recover a slope {$m = 0.0 \pm 0.5$} and an intercept $b = 0.2 \pm 0.2$. A plot of this fit is shown in Figure~\ref{fig:CMF-scatter}.

We also perform an independent analysis using orthogonal distance regression (ODR) with \texttt{SciPy.odr} \citep{VirtanenGommers2020}. As the CMF$_{planet}$ distributions are not Gaussian for many of the stars (being truncated at $0$), we must repeat the same bootstrapping as before. We recover a slope of {$m = 1.9 \pm 7.5$}, and an intercept of {$b = -0.4 \pm 2.2$}. We note that the slope obtained from ODR is highly unconstrained, likely due to our small sample size, high uncertainties in both CMF estimates, and lack of planets with very high or very low CMF$_{star}$ values to leverage the linear fit. As such, we only compare our OLS fit results to the results from other studies. A comparison of the best-fit slopes, as well as the best-fit value of CMF$_{planet}$ (or a proxy thereof) at a solar CMF$_{star} = 0.3$ (the value calculated by \texttt{exopie} at solar abundances) is given in Table \ref{table:CMF-slopes}.

\begin{table*}[] 
\centering
\caption{Comparison to other papers}
\begin{tabular}{c|c|c|c|c}
\hline \hline
Paper & Values Fit & Spectral Type Range & Slope & CMF$_{planet}$ (or similar) at Solar CMF \\ \hline
This paper & CMF$_{planet}$ vs. CMF$_{star}$ & M & {$0.0 \pm 0.5$} & {$0.19 \pm 0.05$} \\
\citet{AdibekyanDorn2021} & Fe$\%_{planet}$ vs. Fe$\%_{star}$ & FGK & $6.3 \pm 1.2$ & $\sim 0.51$ \\
\citet{LiuNi2023} & Fe$\%_{planet}$ vs. Fe$\%_{star}$ & FGKM & $10.80 \pm 3.56$ & $\sim 0.43$ \\
\citet{AdibekyanDeal2024} & Fe$\%_{planet}$ vs. Fe$\%_{star}$ & FGK & $5.85 \pm 1.07$ & $\sim 0.55$ \\
\citet{BrinkmanPolanski2024} & CMF$_{planet}$ vs. CMF$_{star}$ & GK & $1.3 \pm 1.0$ & $0.34 \pm 0.05$ \\
\citet{BrinkmanWeiss2025} & CMF$_{planet}$ vs. CMF$_{star}$ & GK & $1.3 \pm 1.0$ & $0.37 \pm 0.05$ \\ \hline\hline
\end{tabular} \label{table:CMF-slopes}
\smallskip
    \footnotesize

    The results of quantifying a relationship between planetary and stellar CMF estimates (or proxies thereof) between this paper and other papers quantifying similar relationships.
\end{table*}

{A one-to-one relationship between CMF$_{planet}$ and CMF$_{star}$ (corresponding to a slope of one) suggests that planets and their host stars have similar compositions. A slope greater than one, meanwhile, suggests that iron-rich stars host planets even further enriched beyond their stellar composition, while a slope of zero would suggest that the iron-to-silicate ratio of the planet is completely uncorrelated with the composition of its host star. Our slope of $0.0 \pm 0.5$, represents a $2\sigma$ departure from the hypothesis that planets and their host stars have equal iron-to-silicate ratios, though the small sample size may make drawing concrete conclusions from this relationship difficult.}

{These results appear to mark a departure from other works performing similar analyses on primarily FGK stars, as well.} The slope of the relationship between CMF$_{planet}$ and CMF$_{star}$ (or proxies thereof) varies significantly across previous analyses {of different samples of planet-hosting stars}. \citet{AdibekyanDorn2021}, \citet{LiuNi2023}, and \citet{AdibekyanDeal2024} identified very steep slopes, much greater than unity, while \citet{BrinkmanPolanski2024} and \citet{BrinkmanWeiss2025} identify slopes of $1.3 \pm 1.0$, thus finding no strong evidence for a relationship between star and planet abundances. We find a similarly weak linear relationship to \citet{BrinkmanPolanski2024} and \citet{BrinkmanWeiss2025}, which is marginally weaker but only at the $1.0\sigma$ level. It should be noted that some planets in the sample presented here do appear in \citet{LiuNi2023}; this work tends to find substantially lower CMF estimates across this shared sample. We chalk this up to the fact that we homogenously reanalyze stellar masses and radii in this sample, and substantially improve the mass precision for many of the shared targets.

We also note that we recover substantially lower CMF$_{planet}$ values than in any of the aforementioned studies. In Table \ref{table:CMF-slopes}, we report the resulting CMF$_{planet}$ values from the various linear fits, evaluated at the solar CMF$_{star}=0.3$ (equivalently, $0.332$ for Fe$\%_{star}$ as calculated in \citet{AdibekyanDorn2021}). The value of CMF$_{planet}$ (or a similar proxy) at solar abundance/CMF is not generally reported, and so for papers other than this one it is estimated from plots in the corresponding papers. We find that our OLS fit predicts {CMF$_{planet}=0.19 \pm 0.05$} at solar CMF$_{star}$. This value is in marked contrast to previous analyses for planets around predominantly FGK stars, all of which predict CMF$_{planet}$ or Fe$\%_{planet}$ values of $0.34-0.55$ at solar CMF$_{star}$. Our result differs from all previous analyses described above by at least $1.5\sigma$, with studies analyzing primarily FGK stars finding substantially higher CMF estimates than the planets around M dwarfs presented in this study. We conclude that close-in super-Earths around M dwarfs show evidence for lower CMFs than around FGK stars, which we interpret as evidence for larger WMFs.

\subsection{Recommending future observing strategies} \label{subsec:disc-future-obs-strategy}

While the CMF$_{planet}$ and CMF$_{star}$ distributions presented in this work are statistically distinct, {future studies studying compositions of rocky planets around M dwarfs on a population level may strive to better characterize these parameters}. This may be accomplished by either improving mass measurement precisions or by increasing the planet sample size. Here we calculate the total observing times required for both approaches to identify which is more efficient at improving the characterization of the CMF$_{planet}$ and CMF$_{star}$ distributions.

\begin{figure}
    \centering
    \includegraphics[width=0.95\linewidth]{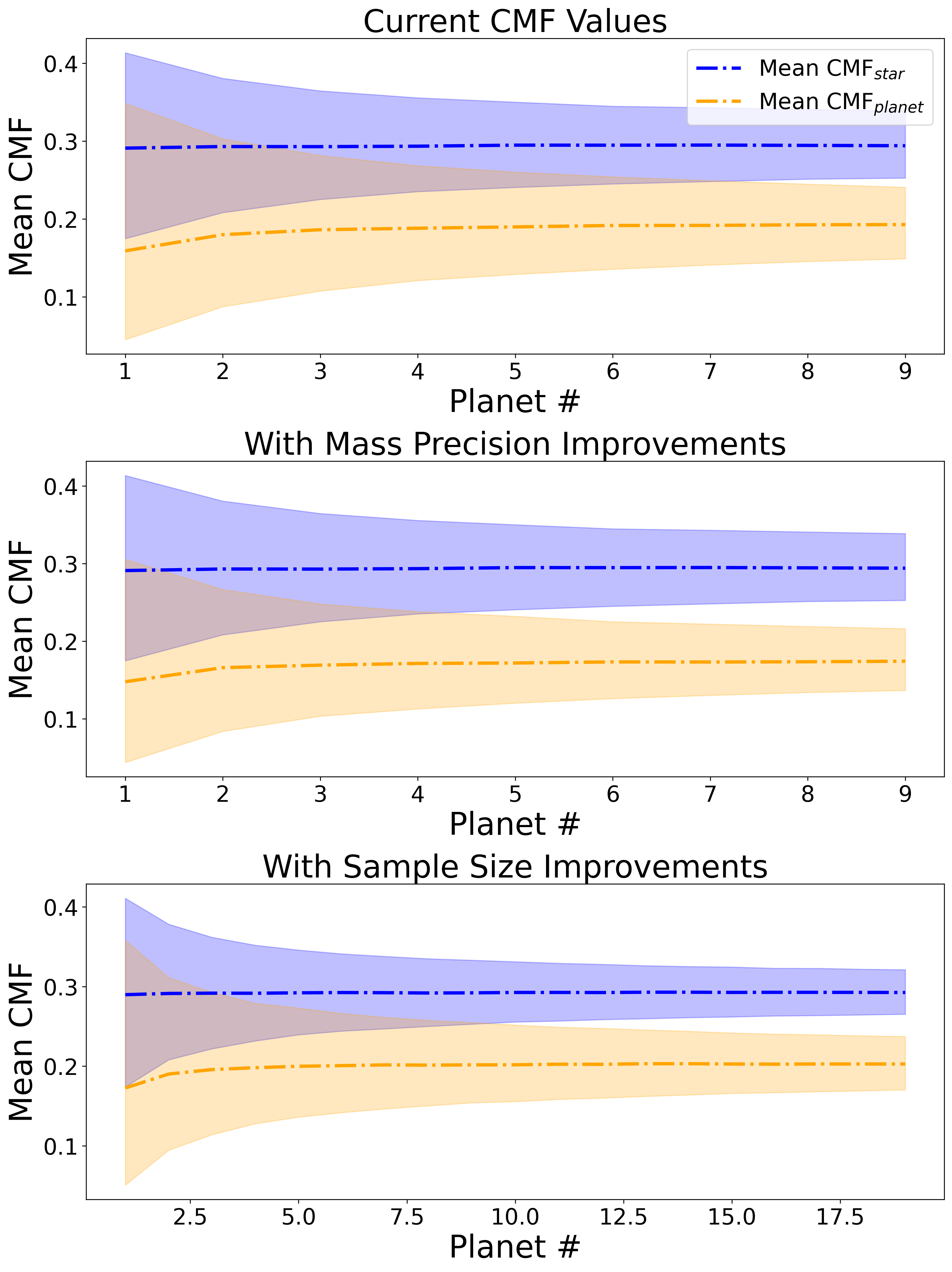}
    \caption{The distributions of the mean CMF$_{planet}$ and CMF$_{star}$ values as a function of the number of planets, for our current sample (top), for an observation campaign focused on improving the masses of planets already in our current sample (middle), and for an observation campaign focused on increasing the size of our sample (bottom). Dashed lines represent the median values of our CMF distributions, while the shaded regions represent the $1\sigma$ upper and lower regions.}
    \label{fig:CMF-improvements}
\end{figure}

We begin by defining a proxy for the distinctness of these distributions: the difference between the posterior of the mean CMF$_{planet}$ and CMF$_{star}$ values in our sample. To get this value, we resample both CMF values of each planet from their posteriors $10^4$ times (c.f., Figure \ref{fig:CMF-hists}) and track the mean CMF as a function of the number of planets sampled. The results of this exercise are shown in the upper panel of Figure~\ref{fig:CMF-improvements}. With our current sample of {nine} planets, we achieve a {$1.6\sigma$} difference between our mean CMF$_{planet}$ and CMF$_{star}$ distributions. We note, however, that while {$1.6\sigma$} may not seem like much, we have already shown the two CMF distributions are indeed statistically different.

Next, we then calculate the improvement in the mean CMF precisions that are expected to be gained from a hypothetical observing campaign aimed at improving the mass measurements precisions of the {nine} targets in our sample to $10\sigma$.
We assume that the resulting CMF$_{planet}$ posteriors have identical means to their current values, but have uncertainties identical to the current uncertainty of the CMF$_{planet}$ of LTT\,3780\,b $(0.15)$. We also assume that the CMF$_{star}$ distributions would remain unchanged. We then repeat the same resampling process to compute the resulting mean CMF$_{planet}$ distribution. The results of this calculations are shown in the middle panel of Figure~\ref{fig:CMF-improvements}. As expected, the assumed $10\sigma$ mass measurements improve the CMF$_{planet}$ precision, which would then be distinguishable from the CMF$_{star}$ distribution at {$2.0\sigma$}, compared to the {$1.6\sigma$} difference achieved with our current sample.

Lastly, we repeat this calculation for a hypothetical observing campaign that increases the size of the planet sample with $\sim 5\sigma$ mass measurements precision (roughly the lowest mass precision in our current sample). We adopt our mean CMF distributions with our current sample size of {nine} and iteratively add new planets assuming that they follow the current CMF$_{planet}$ distribution. For each added planet, we set its CMF$_{star}$ distribution to that system's CMF$_{star}$ distribution from Table~\ref{table:planet-params}. We also assume these new planets have CMF$_{planet}$ posteriors sampled from a truncated Gaussian distribution, with a mean equal to the CMF$_{planet}$ mean of the same planet but with an uncertainty of $0.22$, as this value is roughly equal to the CMF$_{planet}$ uncertainties of GJ\,357\,b and L\,98-59\,b, the two targets in our sample with $\sim5\sigma$ mass precisions. We then repeat the same resampling process as before to compute the mean CMF distributions as a function of the number of planets. The results of this exercise are shown in the lower panel of Figure \ref{fig:CMF-improvements}. We find that to match the {$2.0\sigma$} difference between our mean CMF$_{planet}$ and CMF$_{star}$ distributions, {ten} new targets with $5\sigma$ masses are needed.

Given these two possible scenarios, here we calculate which strategy requires fewer RV observations and can therefore be completed more efficiently. We use the prescription in \citet{CloutierDoyon2018} to calculate the precision in planetary RV semiamplitudes given the precision of RV observations (including RV jitter from residual systematics and stellar activity) \footnote{The use of 3 as the coefficient instead of 2 for this equation is at the recommendation of the author of \citet{CloutierDoyon2018}, who advises it as an approximation in the case of moderate stellar variability \citep[e.g.,][]{BurtZellem2025}.}:

\begin{equation}
N_{RV} = 3\left(\frac{\sigma_\textrm{eff}}{\sigma_\textrm{K}}\right)^2
\end{equation}

where $\sigma_\textrm{eff}$ is the effective RV uncertainty of the observations and $\sigma_\textrm{K}$ is the goal semiamplitude uncertainty. We assume fiducial values of $\sigma_{\rm RV} = 2.5$ m/s (approximately the RV uncertainty for GJ\,1132, the target in our sample with the lowest NIRPS RV precision), and assume a fixed semiamplitude of $K = 3$ m/s for each new planet. We further assume that any observing campaign focused on improving mass precisions of the existing sample would seek to improve the mass precision of all targets in this sample that do not already have a $\sim10\sigma$ mass detection (i.e., \hbox{GJ\,357\,b}, \hbox{HD\,260655\,b}, \hbox{L\,98-59\,b}, and \hbox{TOI-270\,b}). With the fiducial input values described above, improving the mass precisions of these four targets to $10\sigma$ would require approximately 510, 410, 5900, and 360 RVs respectively. We then repeat this calculation for the alternative observing strategy of targeting {ten} planets with $5\sigma$ mass precisions. We find that $5\sigma$ masses can be achieved with roughly 52 RVs for each planet, meaning increasing the size of this sample by nine would take roughly {520} total RVs, far smaller than the total number of RVs necessary to improve the mass precision of even three of the four aforementioned targets (1280 RVs). It is clear that future observations aimed at improving the statistical significance of the conclusions drawn in this paper would be far better served by increasing the sample size of hot super-Earths around M dwarfs with precisions of roughly $5\sigma$ rather than conducting intensive RV follow-up to improve the mass precision of existing targets to $\sim10\sigma$.

\section{Conclusions} \label{sec:conclusions}

This paper has presented the first results from the CMF subprogramme within the NIRPS GTO program. We presented data from the NIRPS and HARPS spectrographs measuring precise RVs of three M dwarfs hosting hot super-Earths -- GJ\,1132, GJ\,1252, and LTT\,3780 -- which substantially improve their mass measurement precisions to $\sim$$10\sigma$. We also present measurements of the stellar abundances of the refractory elements Mg, Si, Fe, and in some cases Ti using spectral synthesis methods. We measure the abundances for the three aforementioned M dwarfs plus five other M dwarfs hosting hot super-Earths (i.e., GJ\,357, HD\,260655, L 98-59, LHS\,1140, and TOI-270). We calculated the core mass fractions of all {nine} planets from their masses and radii (CMF$_{planet}$) as well as the equivalent CMFs inferred from their host stars' elemental abundances (CMF$_{star}$). Our main findings are summarized below.

\begin{itemize}
    \item We recover planetary CMF$_{planet}$ values to a precision of $12-15\%$ for our three targets with intensive RV follow-up with NIRPS and HARPS. We recover a typical CMF$_{planet}$ precision of $\sim20\%$ for the five remaining targets. We also recover CMF$_{star}$ values implied by our host stellar abundances to a typical precision of $\sim$$10\%$.
    \item {Eight of our nine} target planets have smaller CMFs than expected from their host star abundances. We show that this result {cannot be attributed to random} chance. We do not see strong evidence that the host star abundances track the planetary CMFs for hot super-Earths around M dwarfs.
    \item The significant discrepancy between the CMF$_{planet}$ and CMF$_{star}$  distributions suggests that these planets have a significant mass fraction of water. This water is likely sequestered in the planetary interior rather than differentiated on the surface, due to the planets' high equilibrium temperatures that are expected to vaporize any surface water.
    \item We calculate the WMFs necessary to produce these discrepancies assuming that the water is either sequestered or fully differentiated. The WMFs needed to reconcile our observed CMF$_{planet}$-CMF$_{star}$ discrepancy is on the order of $\sim 1\%$ regardless of whether the water is sequestered or differentiated.
    \item None of our planets show evidence for having a larger CMF$_{planet}$ than the CMF$_{star}$ expected from the stellar abundances. This lack of super-Mercuries in our sample suggests that {these planets may be uncommon around M dwarfs, though the small sample size limits the inferences that can be made about their occurrence rates or formation mechanisms}.
    \item We recommend that future observing campaigns seeking to validate the results of our study focus on increasing the sample size of planets with $\sim 5\sigma$ mass precisions rather than trying to obtain ultra-precise mass measurements of the {nine} planets in our current sample.
\end{itemize}

\begin{acknowledgements}

This publication makes use of The Data \& Analysis Center for Exoplanets (DACE), which is a facility based at the University of Geneva dedicated to extrasolar planets data visualization, exchange, and analysis.\\
DW thanks Alison Sills for their scientific discussions that improved the quality of the statistical analysis, {and Bennett Skinner for their scientific discussions regarding interior structure models}.\\
RC acknowledges the support of the Natural Sciences and Engineering Research Council of Canada (NSERC).\\
CC, AL, \'EA, FBa, BB, NJC, RD, LMa, RA, LB, AD-B, LD, PLam, OL, LMo, TV \& JPW  acknowledge the financial support of the FRQ-NT through the Centre de recherche en astrophysique du Qu\'ebec as well as the support from the Trottier Family Foundation and the Trottier Institute for Research on Exoplanets.\\
AL  acknowledges support from the Fonds de recherche du Qu\'ebec (FRQ) - Secteur Nature et technologies under file \#349961.\\
AC, XB, XDe, TF \& VY  acknowledge funding from the French ANR under contract number ANR\-24\-CE49\-3397 (ORVET), and the French National Research Agency in the framework of the Investissements d'Avenir program (ANR-15-IDEX-02), through the funding of the ``Origin of Life" project of the Grenoble-Alpes University.\\
\'EA, FBa, RD \& LMa  acknowledges support from Canada Foundation for Innovation (CFI) program, the Universit\'e de Montr\'eal and Universit\'e Laval, the Canada Economic Development (CED) program and the Ministere of Economy, Innovation and Energy (MEIE).\\
SCB, EC, NCS, ED-M, JGd \& ARCS  acknowledge the support from FCT - Funda\c{c}\~ao para a Ci\^encia e a Tecnologia through national funds by these grants: UIDB/04434/2020, UIDP/04434/2020.\\
SCB   acknowledges the support from Funda\c{c}\~ao para a Ci\^encia e Tecnologia (FCT) in the form of a work contract through the Scientific Employment Incentive program with reference 2023.06687.CEECIND and DOI \href{https://doi.org/10.54499/2023.06687.CEECIND/CP2839/CT0002}{10.54499/2023.06687.CEECIND/CP2839/CT0002.}\\
NBC  acknowledges support from an NSERC Discovery Grant, a Canada Research Chair, and an Arthur B. McDonald Fellowship, and thanks the Trottier Space Institute for its financial support and dynamic intellectual environment.\\
XDu  acknowledges the support from the European Research Council (ERC) under the European Union’s Horizon 2020 research and innovation programme (grant agreement SCORE No 851555) and from the Swiss National Science Foundation under the grant SPECTRE (No 200021\_215200).\\
This work has been carried out within the framework of the NCCR PlanetS supported by the Swiss National Science Foundation under grants 51NF40\_182901 and 51NF40\_205606.\\
DE  acknowledge support from the Swiss National Science Foundation for project 200021\_200726. The authors acknowledge the financial support of the SNSF.\\
JIGH, ASM, RRe, NN \& AKS  acknowledge financial support from the Spanish Ministry of Science, Innovation and Universities (MICIU) projects PID2020-117493GB-I00 and PID2023-149982NB-I00.\\
The Board of Observational and Instrumental Astronomy (NAOS) at the Federal University of Rio Grande do Norte's research activities are supported by continuous grants from the Brazilian funding agency CNPq. This study was partially funded by the Coordena\c{c}\~ao de Aperfei\c{c}oamento de Pessoal de N\'ivel Superior—Brasil (CAPES) — Finance Code 001 and the CAPES-Print program.\\
ICL  acknowledges CNPq research fellowships (Grant No. 313103/2022-4).\\
BLCM  acknowledge CAPES postdoctoral fellowships.\\
BLCM  acknowledges CNPq research fellowships (Grant No. 305804/2022-7).\\
JRM  acknowledges CNPq research fellowships (Grant No. 308928/2019-9).\\
CMo  acknowledges the funding from the Swiss National Science Foundation under grant 200021\_204847 “PlanetsInTime”.\\
Co-funded by the European Union (ERC, FIERCE, 101052347). Views and opinions expressed are however those of the author(s) only and do not necessarily reflect those of the European Union or the European Research Council. Neither the European Union nor the granting authority can be held responsible for them.\\
GAW is supported by a Discovery Grant from the Natural Sciences and Engineering Research Council (NSERC) of Canada.\\
0\\
RA  acknowledges the Swiss National Science Foundation (SNSF) support under the Post-Doc Mobility grant P500PT\_222212 and the support of the Institut Trottier de Recherche sur les Exoplan\`etes (IREx).\\
LB  acknowledges the support of the Natural Sciences and Engineering Research Council of Canada (NSERC).\\
This project has received funding from the European Research Council (ERC) under the European Union's Horizon 2020 research and innovation programme (project {\sc Spice Dune}, grant agreement No 947634). This material reflects only the authors' views and the Commission is not liable for any use that may be made of the information contained therein.\\
LD  acknowledges the support of the Natural Sciences and Engineering Research Council of Canada (NSERC) and from the Fonds de recherche du Qu\'ebec (FRQ) - Secteur Nature et technologies.\\
ED-M  further acknowledges the support from FCT through Stimulus FCT contract 2021.01294.CEECIND. ED-M  acknowledges the support by the Ram\'on y Cajal contract RyC2022-035854-I funded by MICIU/AEI/10.13039/501100011033 and by ESF+.\\
CD acknowledges support from the Swiss National Science Foundation under grant TMSGI2\_211313. This work has been carried out in parts within the framework of the NCCR \verb|PlanetS| supported by the Swiss National Science Foundation under grant 51NF40\_205606.\\
FG  acknowledges support from the Fonds de recherche du Qu\'ebec (FRQ) - Secteur Nature et technologies under file \#350366.\\
This work was supported by grants from eSSENCE (grant number eSSENCE@LU 9:3), the Swedish National Research Council (project number 2023 05307), The Crafoord foundation and the Royal Physiographic Society of Lund, through The Fund of the Walter Gyllenberg Foundation."\\
LMo  acknowledges the support of the Natural Sciences and Engineering Research Council of Canada (NSERC), [funding reference number 589653].\\
KAM  acknowledges support from the Swiss National Science Foundation (SNSF) under the Postdoc Mobility grant P500PT\_230225.\\
NN  acknowledges financial support by Light Bridges S.L, Las Palmas de Gran Canaria.\\
NN acknowledges funding from Light Bridges for the Doctoral Thesis "Habitable Earth-like planets with ESPRESSO and NIRPS", in cooperation with the Instituto de Astrof\'isica de Canarias, and the use of Indefeasible Computer Rights (ICR) being commissioned at the ASTRO POC project in the Island of Tenerife, Canary Islands (Spain). The ICR-ASTRONOMY used for his research was provided by Light Bridges in cooperation with Hewlett Packard Enterprise (HPE).\\
CP  acknowledges support from the NSERC Vanier scholarship, and the Trottier Family Foundation. CP  also acknowledges support from the E. Margaret Burbidge Prize Postdoctoral Fellowship from the Brinson Foundation.\\
ARCS  acknowledges the support from Funda\c{c}ao para a Ci\^encia e a Tecnologia (FCT) through the fellowship 2021.07856.BD.\\
AKS  acknowledges financial support from La Caixa Foundation (ID 100010434) under the grant LCF/BQ/DI23/11990071.\\
TV  acknowledges support from the Fonds de recherche du Qu\'ebec (FRQ) - Secteur Nature et technologies under file \#320056.

\end{acknowledgements}

\bibliography{bibliography}

\clearpage
\onecolumn

\appendix

\section{Stellar and planetary parameters}\label{appendix:stell-plan-parameters}

\begin{table}[!h]
\centering
    \begin{adjustbox}{angle=90}
    \begin{threeparttable}
    \caption{Reported stellar parameters}
    \begin{tabularx}{1.25\linewidth}{C C C C C C C C C}
    \hline\hline
Parameter & GJ 1132 & GJ 1252 & LTT 3780 & GJ 357 & HD\,260655 & L 98-59 & LHS\,1140 & TOI-270 \\
\hline
TOI & 667 & 1078 & 732 & 562 & 4599 & 175 & 256 & 270 \\
Mass ($M_\odot$) & $0.198 \pm 0.005$ & $0.381 \pm 0.008$ & $0.363 \pm 0.008$ & $0.346 \pm 0.007$ & $0.463 \pm 0.010$ & $0.293 \pm 0.006$ & $0.184 \pm 0.004$ & $0.362 \pm 0.008$ \\
Radius ($R_\odot$) & $0.226 \pm 0.007$ & $0.391 \pm 0.012$ & $0.375 \pm 0.011$ & $0.360 \pm 0.011$ & $0.465 \pm 0.014$ & $0.314 \pm 0.009$ & $0.215 \pm 0.007$ & $0.374 \pm 0.011$ \\
$T_\textrm{eff}$ (K) & $3229_{-62}^{+78}$\tnote{3} & $3458 \pm 137$\tnote{4} & $3358 \pm 92$\tnote{5} & $3505 \pm 51$\tnote{6} & $3803 \pm 50$\tnote{7} & $3415 \pm 135$\tnote{8} & $3096 \pm 48$\tnote{9} & $3506 \pm 70$\tnote{10} \\
Sp. T & M4.5V & M3V & M3.5V & M2.5V & M0V & M3V & M4.5V & M3V \\
log g & $5.02 \pm 0.03$ & $4.83 \pm 0.03$ & $4.85 \pm 0.03$ & $4.86 \pm 0.03$ & $4.77 \pm 0.03$ & $4.91 \pm 0.03$ & $5.04 \pm 0.03$ & $4.85 \pm 0.03$ \\
$v_\textrm{mic}$ (km/s) & $[0, 2]$ & $[0, 1]$ & $[0, 1]$ & $[0, 1]$ & $[0, 1.5]$ & $[0, 1.5]$ & $[0, 1]$ & $[0, 1]$ \\
$v_\textrm{mac}$ (km/s) & $[5.5, 7.5]$ & $5.65 \pm 1.98$ & $5.54 \pm 2.54$ & $5.48 \pm 1.86$ & $4.33 \pm 1.10$ & $5.43 \pm 2.36$ & $[6,8]$ & $5.33 \pm 1.73$ \\
{[Fe/H]} & $-0.18 \pm 0.15$ & $-0.18 \pm 0.13$ & $0.06 \pm 0.14$ & $-0.12 \pm 0.12$ & $-0.48 \pm 0.06$ & $-0.29 \pm 0.15$ & $-0.25 \pm 0.11$ & $-0.14 \pm 0.12$ \\
{[Mg/H]} & --\tnote{1} & $-0.04 \pm 0.24$ & $0.07 \pm 0.16$ & $-0.27 \pm 0.12$ & $-0.32 \pm 0.09$ & $-0.34 \pm 0.21$ & --\tnote{1} & $-0.31 \pm 0.13$ \\
{[Si/H]} & --\tnote{1} & $-0.00 \pm 0.39$ & $0.20 \pm 0.23$ & $-0.26 \pm 0.16$ & $-0.57 \pm 0.11$ & $-0.27 \pm 0.28$ & --\tnote{1} & $-0.20 \pm 0.19$ \\
{[Ti/H]} & $-0.19 \pm 0.17$ & -- & -- & -- & -- & -- & $-0.30 \pm 0.12$ & -- \\
{[M/H]} & $-0.17 \pm 0.10$ & $-0.02_{-0.16}^{+0.19}$ & $0.14_{-0.11}^{+0.12}$ & $-0.20 \pm 0.08$ & $-0.43 \pm 0.06$ & $-0.27_{-0.13}^{+0.14}$ & $-0.28 \pm 0.07$ & $-0.20 \pm 0.09$ \\
{Fe/Mg} & $0.81_{-0.33}^{+0.56}$ \tnote{2} & $0.57_{-0.27}^{+0.51}$ & $0.77_{-0.30}^{+0.50}$ & $1.12_{-0.37}^{+0.56}$ & $0.55_{-0.14}^{+0.19}$ & $0.89_{-0.40}^{+0.72}$ & $0.90_{-0.29}^{+0.44}$ \tnote{2} & $1.17_{-0.40}^{+0.62}$ \\
{Mg/Si} & -- & $1.16_{-0.76}^{+2.15}$ & $0.92_{-0.44}^{+0.84}$ & $1.22_{-0.46}^{+0.74}$ & $2.19_{-0.66}^{+0.94}$ & $1.07_{-0.59}^{+1.33}$ & -- & $0.94_{-0.40}^{+0.69}$ \\
{[$\alpha$/Fe]} & $0.00 \pm 0.19$ & $0.20_{-0.26}^{+0.27}$ & $0.10 \pm 0.20$ & $-0.14 \pm 0.16$ & $0.07 \pm 0.10$ & $0.01 \pm 0.23$ & $-0.05 \pm 0.14$ & $-0.10_{-0.16}^{+0.17}$ \\

\hline

    \hline
    \end{tabularx}\label{table:stellar-params}
    
    \smallskip
    \footnotesize

    This table details the stellar parameters and abundances derived in this paper. Stellar masses and radii are derived from relations in \citet{MannFeiden2015} and \citet{MannDupuy2019}, using K-band magnitudes from 2MASS and parallaxes from Gaia DR3. Surface gravities are derived from calculated masses and radii. The {[M/H]} values reported here are the final values derived from stellar abundances in this paper; the initial {[M/H]} values used in the stellar abundance calculations described in Section \ref{subsec:stellchar-abund} are given in Section \ref{subsec:stellchar-params}. As we do not calculate effective temperature, however, the values reported in Section \ref{subsec:stellchar-params} are repeated here. For reference, solar values of Fe/Mg and Mg/Si are $0.80_{-0.10}^{+0.11}$ and $1.23_{-0.13}^{+0.15}$ respectively \citep{AsplundGrevesse2009}. All of the microturbulences and some of the macroturbulences calculated have unreasonably wide uncertainties or physically implausible parameter ranges when fit as described in Section \ref{subsec:stellchar-abund}; as such, we estimate by eye these parameter ranges and report them instead.

    \begin{tablenotes}
    \item[1] The [Mg/H] and [Si/H] abundances for GJ 1132 and LHS\,1140 are not considered reliable, due to the low number of spectral lines used and their poor quality. See Section \ref{subsec:stellchar-abund} for more details; the unreliable abundances are reported in Table \ref{table:abundances-appendix} for completeness.
    \item[2] The Fe/Mg values calculated for GJ 1132 and LHS\,1140 are calculated by assuming [Mg/H] is equal to the calculated [Ti/H] value; see Section \ref{subsec:stellchar-abund} for more details.
    \item[3] {From \citet{XueBean2024}.}
    \item[4] {From \citet{CrossfieldMalik2022}.}
    \item[5] {From \citet{BonfantiBrady2024}.}
    \item[6] {From \citet{SchweitzerPassegger2019}.}
    \item[7] {From \citet{MarfilTabernero2021}.}
    \item[8] {From \citet{DemangeonZapateroOsorio2021}.}
    \item[9] {From \citet{CadieuxPlotnykov2024}.}
    \item[10] {From \citet{VanEylenAstudillo-Defru2021}.}
    
    \end{tablenotes}

    \end{threeparttable}
    \end{adjustbox}
\end{table}

\begin{table}
\centering
    \begin{threeparttable}
    \caption{Reported planetary parameters}
    
    \begin{tabularx}{\linewidth}{C C C C C C}
    \hline\hline
Parameter & GJ\,1132\,b & GJ\,1252\,b & LTT\,3780\,b & GJ\,357\,b & HD\,260655\,b \\
\hline
Period (days) & $1.63$ & $0.52$ & $0.77$ & $3.93$ & $2.77$ \\
Mass ($M_\oplus$) & $1.69 \pm 0.15$ & $1.54 \pm 0.18$ & $2.34 \pm 0.10$ & $1.83 \pm 0.30$ & $2.19 \pm 0.35$ \\
Radius ($R_\oplus$) & $1.22 \pm 0.04$ & $1.19 \pm 0.05$ & $1.31 \pm 0.05$ & $1.21 \pm 0.05$ & $1.31_{-0.04}^{+0.05}$ \\
T$_{eq}$ (K) & $591 \pm 11$ & $1090 \pm 46$ & $916 \pm 29$ & $548 \pm 11$ & $724 \pm 14$ \\
CMF$_{planet}$ & $<0.39$ & $<0.48$ & $0.19_{-0.12}^{+0.15}$ & $0.27_{-0.18}^{+0.20}$ & $<0.46$ \\
CMF$_{star}$ & $0.29_{-0.11}^{+0.12}$ & $0.20_{-0.12}^{+0.14}$ & $0.23_{-0.11}^{+0.12}$ & $0.38 \pm 0.11$ & $0.28 \pm 0.06$ \\ \hline \hline
Parameter & L\,98-59\,b & L\,98-59\,c & LHS\,1140\,c & TOI-270\,b & \\ \hline
Period (days) & $2.25$ & $3.69$ & $3.78$ & $3.36$ & \\
Mass ($M_\oplus$) & $0.45 \pm 0.11$ & $1.98\pm0.13$ & $1.89 \pm 0.06$ & $1.49 \pm 0.25$ & \\
Radius ($R_\oplus$) & $0.83 \pm 0.03$ & $1.32\pm 0.04$ & $1.25 \pm 0.04$ & $1.28 \pm 0.04$ & \\
T$_{eq}$ (K) & $617 \pm 26$ & $524 \pm 22$ & $421 \pm 9$ & $585 \pm 15$ & \\
CMF$_{planet}$ & $0.29_{-0.19}^{+0.22}$ & $<0.26$ & {$0.14_{-0.10}^{+0.13}$} & $<0.29$ & \\
CMF$_{star}$ & $0.29_{-0.13}^{+0.14}$ & $0.29_{-0.13}^{+0.14}$ & $0.32 \pm 0.09$ & $0.34_{-0.11}^{+0.12}$ & \\ 
\hline
    \end{tabularx}\label{table:planet-params}

    \smallskip
    \footnotesize

    This table details derived masses, radii, and equilibrium temperatures, as well as previously reported periods of our planetary sample, along with the CMFs calculated for each planet in our sample (both as derived from the planet masses and radii and as inferred from stellar abundances.) 95th percentile upper limits on the CMF estimates are reported instead of error bars where the CMF posterior is consistent with 0 {(i.e., the mode of the CMF posterior distribution is at zero)}. Equilibrium temperatures are reported as if Bond albedo is 0.
    \end{threeparttable}
\end{table}

\section{RV plots}\label{appendix:rv-plots}

\begin{table*}[!h]
\caption{GJ\,1132 RV analysis priors and parameters}
\centering
    \begin{threeparttable}
    \begin{tabularx}{\linewidth}{C C C C}
    \hline\hline
    Parameter & Units & Prior & Posterior \\
    \hline
        $\gamma_{\textrm{HARPS,a}}$ & $m/s$ & $\mathcal{N}(\mu_\textrm{HARPS,a},\sigma_\textrm{HARPS,a})$\tnote{1} & $35078.6 \pm 0.5$ \\
        $\gamma_{\textrm{HARPS,s}}$ & $m/s$ & $\mathcal{N}(\mu_\textrm{HARPS,s},\sigma_\textrm{HARPS,s})$\tnote{1} & $34771.8 \pm 1.3$ \\
        $\gamma_{\textrm{NIRPS}}$ & $m/s$ & $\mathcal{N}(\mu_\textrm{NIRPS},\sigma_\textrm{NIRPS})$\tnote{1} & $35023.1_{-2.8}^{+2.9}$ \\
        $\rho_{GP}$ & days & $\mathcal{SN}(122.3,5.0, 6.0)\tnote{2}$ & $125.6_{-5.9}^{+6.4}$ \\
        $\log \tau$ & days & $\mathcal{U}(\log{(2 P_{GP})}, \log{(100 P_{GP})})$ & $5.61_{-0.09}^{+0.20}$ \\
        $\log \sigma_{\textrm{HARPS,a}}$ & $m^2/s^2$ & $\mathcal{U}(-5, 5)$ & $1.2 \pm 0.3$ \\
        $\log \sigma_{\textrm{HARPS,s}}$ & $m^2/s^2$ & $\mathcal{U}(-5, 5)$ & $2.1 \pm 0.4$ \\
        $\log \sigma_{\textrm{NIRPS}}$ & $m^2/s^2$ & $\mathcal{U}(-5, 5)$ & $3.2 \pm 0.3$ \\
        $\log K_b$ & $m/s$ & $\mathcal{U}(-5, 5)$ & $1.02 \pm 0.09$ \\
        $K_b$ & $m/s$ & --\tnote{3} & $2.75 \pm 0.24$ \\
        $P_b$ & days & $\mathcal{N}(1.628931,0.000027)$ \tnote{2} & $1.62893 \pm 0.00001$ \\
        $t_{0,b}$ & BJD-2450000 & $\mathcal{N}(7184.55786, 0.00031)$ \tnote{2} & $7184.5579 \pm 0.0003$ \\
        $\log K_c$ & $m/s$ & $\mathcal{U}(-5, 5)$ & $0.99_{-0.10}^{+0.09}$ \\
        $K_c$ & $m/s$ & --\tnote{3} & $2.69 \pm 0.25$ \\
        $P_c$ & days & $\mathcal{N}(8.929,0.010)$ \tnote{2} & $8.9290 \pm 0.0008$ \\
        $t_{0,c}$ & BJD-2450000 & $\mathcal{N}(7506.02,0.34)$ \tnote{2} & $7505.94 \pm 0.15$ \\
        $h_c$ & -- & $\mathcal{U}(-1, 1)$ & $-0.05_{-0.20}^{+0.21}$ \\
        $k_c$ & -- & $\mathcal{U}(-1, 1)$ & $-0.05_{-0.20}^{+0.22}$ \\
        $e_c$ & -- & --\tnote{3} & $<0.19$ \\
        $M_b$ & $M_\oplus$ & --\tnote{3} & $1.69 \pm 0.15$ \\
        $M_c$ & $M_\oplus$ & --\tnote{3} & $2.91 \pm 0.27$ \\
    \hline
    \end{tabularx}
    
    \smallskip
    \footnotesize

    The priors and posteriors of the planetary and GP parameters for the GJ\,1132 system. Note that GJ\,1132\,b was restricted to have an eccentricity of $0$ thanks to its short period; GJ\,1132 c's eccentricity was allowed to float. Note that the masses reported is a minimum mass in the case of GJ\,1132 c, as it does not transit. The \texttt{HARPS,a} and \texttt{HARPS,s} subscripts refers to the archival HARPS data and the HARPS data taken simultaneously with NIRPS respectively. $\mathcal{SN}$ refers to a split-normal distribution, parameterized here as $\mathcal{SN}(\mu,\sigma_-,\sigma_+)$.
    
    \begin{tablenotes}
    \item[1] Normal distribution, based on the mean and standard deviation of the data points fit from this instrument.
    \item[2] Obtained from \citet{BonfilsAlmenara2018}.
    \item[3] Derived from other parameters.
    
    \end{tablenotes}
    \end{threeparttable} \label{table:toi667-rv-table}
\end{table*}

\begin{figure*}[!h]
    \centering
    \includegraphics[width=\linewidth]{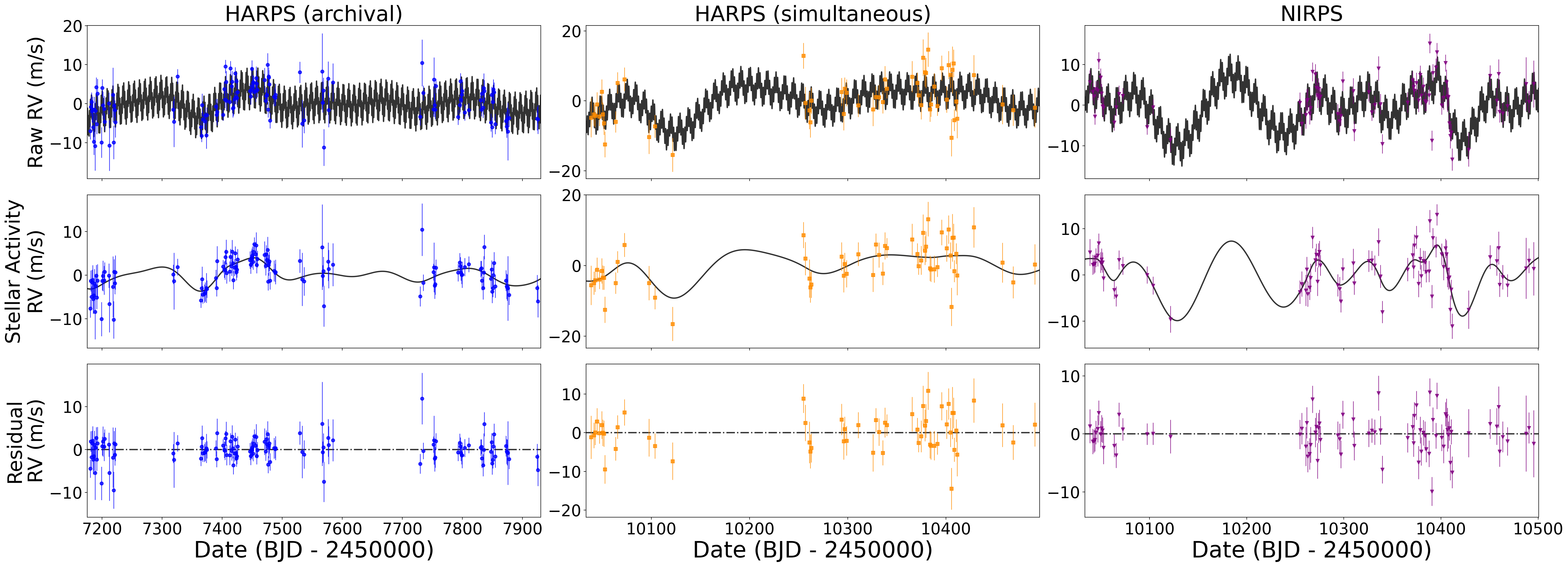}
    \caption{The RV time series used in the GJ\,1132 RV modeling. The top row shows the mean-subtracted RV-time series for each instrument, with the maximum a-posteriori RV model shown in black, while the middle row shows just the stellar activity component of the RV time series, with the corresponding best-fit stellar activity signal shown in black. The residuals between the best-fit model and each time series are shown in the bottom row.}
    \label{fig:gj1132-component}
\end{figure*}

\begin{table*}[!h]
\caption{GJ\,1252 RV analysis priors and parameters}
\centering
    \begin{threeparttable}
    \begin{tabularx}{\linewidth}{C C C C C}
    \hline\hline
    Parameter & Units & Prior & Posterior (1-planet) & Posterior (2-planet) \\
    \hline
    $\gamma_\textrm{HARPS,a}$ & $m/s$ & $\mathcal{N}(\mu_\textrm{HARPS,a},\sigma_\textrm{HARPS,a})$\tnote{1} & $-5.5 \pm 2.5$ & $-5.4 \pm 0.7$ \\
    $\gamma_\textrm{HARPS,s}$ & $m/s$ & $\mathcal{N}(\mu_\textrm{HARPS,s},\sigma_\textrm{HARPS,s})$\tnote{1} & $7210.6 \pm 0.7$ & $7210.6_{-0.7}^{+0.6}$ \\
    $\gamma_\textrm{NIRPS}$ & $m/s$ & $\mathcal{N}(\mu_\textrm{NIRPS},\sigma_\textrm{NIRPS})$\tnote{1} & $7362.0 \pm 2.4$ & $7362.4 \pm 2.3$ \\
    $\rho_\textrm{GP}$ & days & $\mathcal{N}(64, 4)$\tnote{2} & $55 \pm 3$ & $55 \pm 3$ \\
    $\log \tau$ & days & $\mathcal{U}(\log{(2 P_{GP})}, \log{(100 P_{GP})})$ & $5.0_{-0.1}^{+0.2}$ & $4.9_{-0.1}^{+0.2}$ \\
    $\log \sigma_\textrm{HARPS,a}$ & $m^2 / s^2$ & $\mathcal{U}(-5, 5)$ & $2.8 \pm 0.4$ & $0.9 \pm 0.5$ \\
    $\log \sigma_\textrm{HARPS,s}$ & $m^2 / s^2$ & $\mathcal{U}(-5, 5)$ & $1.9 \pm 0.2$ & $1.9 \pm 0.2$ \\
    $\log \sigma_\textrm{NIRPS}$ & $m^2 / s^2$ & $\mathcal{U}(-5, 5)$ & $3.5 \pm 0.3$ & $3.4 \pm 0.3$ \\
    $\log K_b$ & $m/s$ & $\mathcal{U}(-5, 5)$ & $0.79_{-0.14}^{+0.13}$ & $0.86_{-0.12}^{+0.11}$ \\
    $K_b$ & $m/s$ &  --\tnote{3} & $2.20 \pm 0.30$ & $2.36 \pm 0.28$ \\
    $P_b$ & days & $\mathcal{N}(0.5182349, 6.3\textrm{e-6})$\tnote{2} & $0.518235 \pm 6\textrm{e-6}$ & $0.518235 \pm 6\textrm{e-6}$ \\
    $t_{0,b}$ & BJD-2450000 & $\mathcal{N}(8668.09739, 3.2\textrm{e-4})$\tnote{2} & $8668.097 \pm 3\textrm{e-4}$ & $8668.097 \pm 3\textrm{e-4}$ \\
    $\log K_c$ & $m/s$ & $\mathcal{U}(-5, 5)$ & -- & $1.04_{-0.15}^{+0.13}$ \\
    $K_c$ & $m/s$ &  --\tnote{4} & -- & $2.84 \pm 0.40$\\
    $P_c$ & days & $\mathcal{N}(17.50, 0.41)$\tnote{2} & -- & $18.41 \pm 0.01$ \\
    $t_{0,c}$ & BJD-2450000 & $\mathcal{N}(8681.4, 3.2)$\tnote{2} & -- & $8677.6_{-0.7}^{+0.6}$ \\
    $h_c$ & -- & $\mathcal{U}(-1, 1)$\tnote{3} & -- & $0.23_{-0.27}^{+0.19}$ \\
    $k_c$ & -- & $\mathcal{U}(-1, 1)$\tnote{3} & -- & $0.27_{-0.28}^{+0.19}$ \\
    $e_c$ & -- & --\tnote{4} & -- & $0.20_{-0.10}^{+0.11}$ \\
    $M_b$ & $M_\oplus$ & --\tnote{4} & $1.43 \pm 0.19$ & $1.54 \pm 0.18$ \\
    $M_c$ & $M_\oplus$ & --\tnote{4} & -- & $6.08 \pm 0.86$ \\
    \hline
    \end{tabularx} 
    
    \smallskip
    \footnotesize

    The priors and posteriors of the planetary and GP parameters for the GJ\,1252 system. Note that GJ\,1252\,b was restricted to have an eccentricity of $0$ thanks to its short period; GJ\,1252\,(c)'s eccentricity was allowed to float. Note that the masses reported is a minimum mass in the case of GJ\,1252\,(c), as it does not transit. The \texttt{HARPS,a} and \texttt{HARPS,s} subscripts refers to the archival HARPS data and the HARPS data taken simultaneously with NIRPS respectively. $\mathcal{SN}$ refers to a split-normal distribution, parameterized here as $\mathcal{SN}(\mu,\sigma_-,\sigma_+)$.
    
    \begin{tablenotes}
    \item[1] Normal distribution, based on the mean and standard deviation of the data points fit from this instrument.
    \item[2] Obtained from \citet{ShporerCollins2020}.
    \item[3] An additional constraint is put on $h, k$ such that $e^2 = h^2 + k^2 < 1$.
    \item[4] Derived from other parameters.
    
    \end{tablenotes}
    \end{threeparttable} \label{table:toi1078-rv-table}
\end{table*}

\begin{figure*}[!h]
    \centering
    \includegraphics[width=\linewidth]{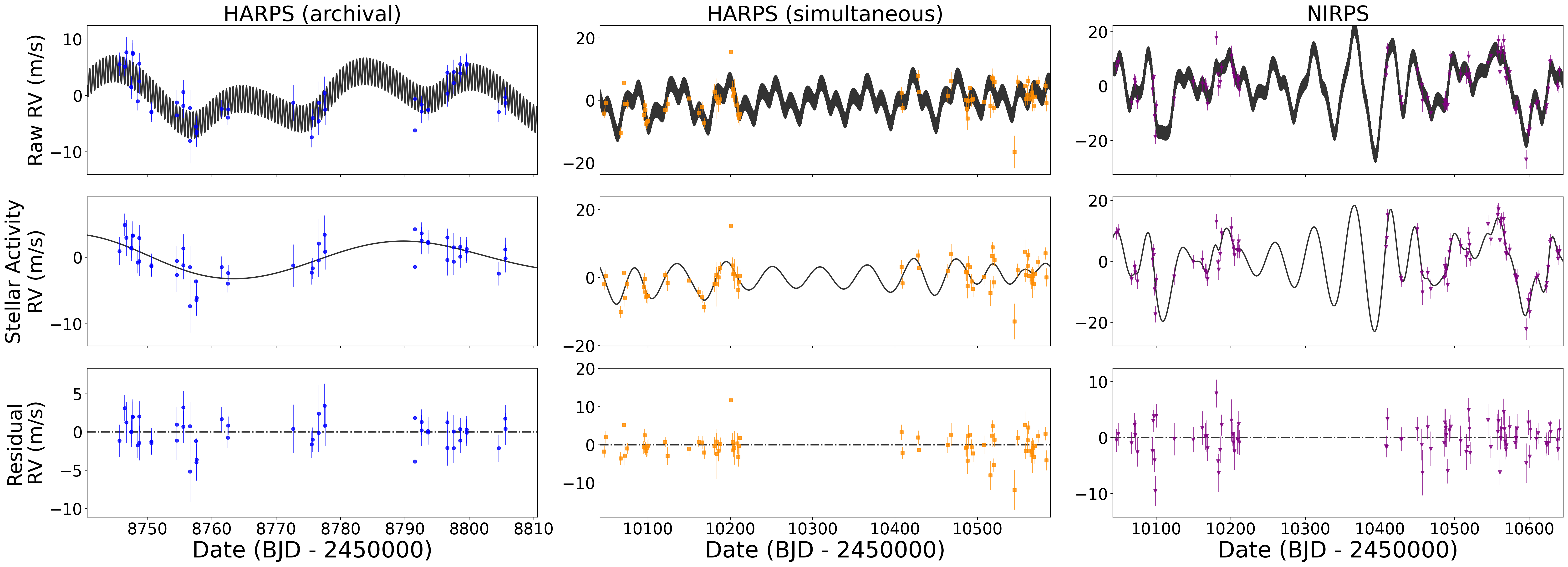}
    \caption{The RV time series used in the GJ\,1252 RV modeling. The top row shows the mean-subtracted RV-time series for each instrument, with the maximum a-posteriori RV model shown in black, while the middle row shows just the stellar activity component of the RV time series, with the corresponding best-fit stellar activity signal shown in black. The residuals between the best-fit model and each time series are shown in the bottom row.}
    \label{fig:gj1252-component}
\end{figure*}

\begin{table*}[!h]
\caption{LTT\,3780 RV analysis priors and parameters}
\centering
    \begin{threeparttable}
    \begin{tabularx}{\linewidth}{C C c c}
    \hline\hline
    Parameter & Units & Prior & Posterior \\
    \hline
        $\gamma_{\textrm{CARMENES}}$ & $m/s$ & $\mathcal{N}(\mu_\textrm{CARMENES},\sigma_\textrm{CARMENES})$\tnote{1} & $-1.2_{-1.2}^{+1.1}$ \\
        $\gamma_{\textrm{HARPS,a}}$ & $m/s$ & $\mathcal{N}(\mu_\textrm{HARPS,a},\sigma_\textrm{HARPS,a})$\tnote{1} & $-0.9 \pm 0.5$ \\
        $\gamma_{\textrm{HARPS-N}}$ & $m/s$ & $\mathcal{N}(\mu_\textrm{HARPS-N},\sigma_\textrm{HARPS-N})$\tnote{1} & $-1.4_{-0.9}^{+0.8}$ \\
        $\gamma_{\textrm{MAROON-X,b}}$ & $m/s$ & $\mathcal{N}(\mu_\textrm{MAROON-X,b},\sigma_\textrm{MAROON-X,b})$\tnote{1} & $1.1 \pm 0.3$ \\
        $\gamma_{\textrm{MAROON-X,r}}$ & $m/s$ & $\mathcal{N}(\mu_\textrm{MAROON-X,r},\sigma_\textrm{MAROON-X,r})$\tnote{1} & $0.7_{-0.3}^{+0.2}$ \\
        $\gamma_{\textrm{NIRPS}}$ & $m/s$ & $\mathcal{N}(\mu_\textrm{NIRPS},\sigma_\textrm{NIRPS})$\tnote{1} & $58.5_{-3.1}^{+3.5}$ \\
        $\rho_{GP}$ & days & $\mathcal{N}(104, 15)\tnote{2}$ & $100.4_{-7.7}^{+9.0}$ \\
        $\log \tau$ & days & $\mathcal{U}(\log{(2 P_{GP})}, \log{(100 P_{GP})})$ & $5.9_{-0.4}^{+0.8}$ \\
        $\log \sigma_{\textrm{CARMENES}}$ & $m^2/s^2$ & $\mathcal{U}(-5, 5)$ & $1.4_{-0.7}^{+0.8}$ \\
        $\log \sigma_{\textrm{HARPS,a}}$ & $m^2/s^2$ & $\mathcal{U}(-5, 5)$ & $0.8_{-0.4}^{+0.5}$ \\
        $\log \sigma_{\textrm{HARPS-N}}$ & $m^2/s^2$ & $\mathcal{U}(-5, 5)$ & $1.3_{-0.6}^{+0.7}$ \\        $\log \sigma_{\textrm{MAROON-X,b}}$ & $m^2/s^2$ & $\mathcal{U}(-5, 5)$ & $-2.3_{-1.8}^{+1.9}$ \\
        $\log \sigma_{\textrm{MAROON-X,r}}$ & $m^2/s^2$ & $\mathcal{U}(-5, 5)$ & $-0.5_{-0.7}^{+0.6}$ \\
        $\log \sigma_{\textrm{NIRPS}}$ & $m^2/s^2$ & $\mathcal{U}(-5, 5)$ & $2.4_{-0.5}^{+0.6}$ \\
        $\log K_b$ & $m/s$ & $\mathcal{U}(-5, 5)$ & $1.18 \pm 0.04$ \\
        $K_b$ & $m/s$ & --\tnote{4} & $3.26 \pm 0.14$ \\
        $P_b$ & days & $\mathcal{SN}(0.76837931,0.00000042,0.00000039)$ \tnote{3} & $0.7683793 \pm 0.0000004$ \\
        $t_{0,b}$ & BJD-2450000 & $\mathcal{SN}(9606.58098,0.00040,0.00032)$ \tnote{3} & $9606.5810 \pm 0.0004$ \\
        $\log K_c$ & $m/s$ & $\mathcal{U}(-5, 5)$ & $1.47 \pm 0.03$ \\
        $K_c$ & $m/s$ & --\tnote{4} & $4.35 \pm 0.13$ \\
        $P_c$ & days & $\mathcal{N}(12.252284,0.000013)$ \tnote{3} & $12.25228 \pm 0.00001$ \\
        $t_{0,c}$ & BJD-2450000 & $\mathcal{SN}(9600.54227,0.00065,00066)$ \tnote{3} & $9600.5423 \pm 0.0007$ \\
        $h_c$ & -- & $\mathcal{U}(-1, 1)$ & $0.12_{-0.11}^{+0.08}$ \\
        $k_c$ & -- & $\mathcal{U}(-1, 1)$ & $0.15_{-0.14}^{+0.10}$ \\
        $e_c$ & -- & --\tnote{4} & $<0.11$ \\
        $M_b$ & $M_\oplus$ & --\tnote{3} & $2.34 \pm 0.10$ \\
        $M_c$ & $M_\oplus$ & --\tnote{3} & $7.89 \pm 0.26$ \\
    \hline
    \end{tabularx} 
    
    \smallskip
    \footnotesize

    The priors and posteriors of the planetary and GP parameters for the LTT\,3780 system. LTT\,3780\,b was restricted to have an eccentricity of $0$ due to its short period; LTT\,3780 c's eccentricity was allowed to float. The \texttt{MAROON-X,r} and \texttt{MAROON-X,b} subscripts refer to the red and blue MAROON-X data respectively; the \texttt{HARPS,a} subscript refers to the archival HARPS data (as opposed to the HARPS data taken simultaneously with NIRPS, which was not included for lack of information content.) $\mathcal{SN}$ refers to a split-normal distribution, parameterized as $\mathcal{SN}(\mu,\sigma_-,\sigma_+)$.
    
    \begin{tablenotes}
    \item[1] Normal distribution, based on the mean and standard deviation of the data points fit from this instrument.
    \item[2] Obtained from \citet{CloutierEastman2020}.
    \item[3] Obtained from \citet{BonfantiBrady2024}.
    \item[4] Derived from other parameters.
    
    \end{tablenotes}
    \end{threeparttable} \label{table:toi732-rv-table}
\end{table*}

\clearpage

\begin{figure*}[!h]
    \centering
    \includegraphics[width=\linewidth]{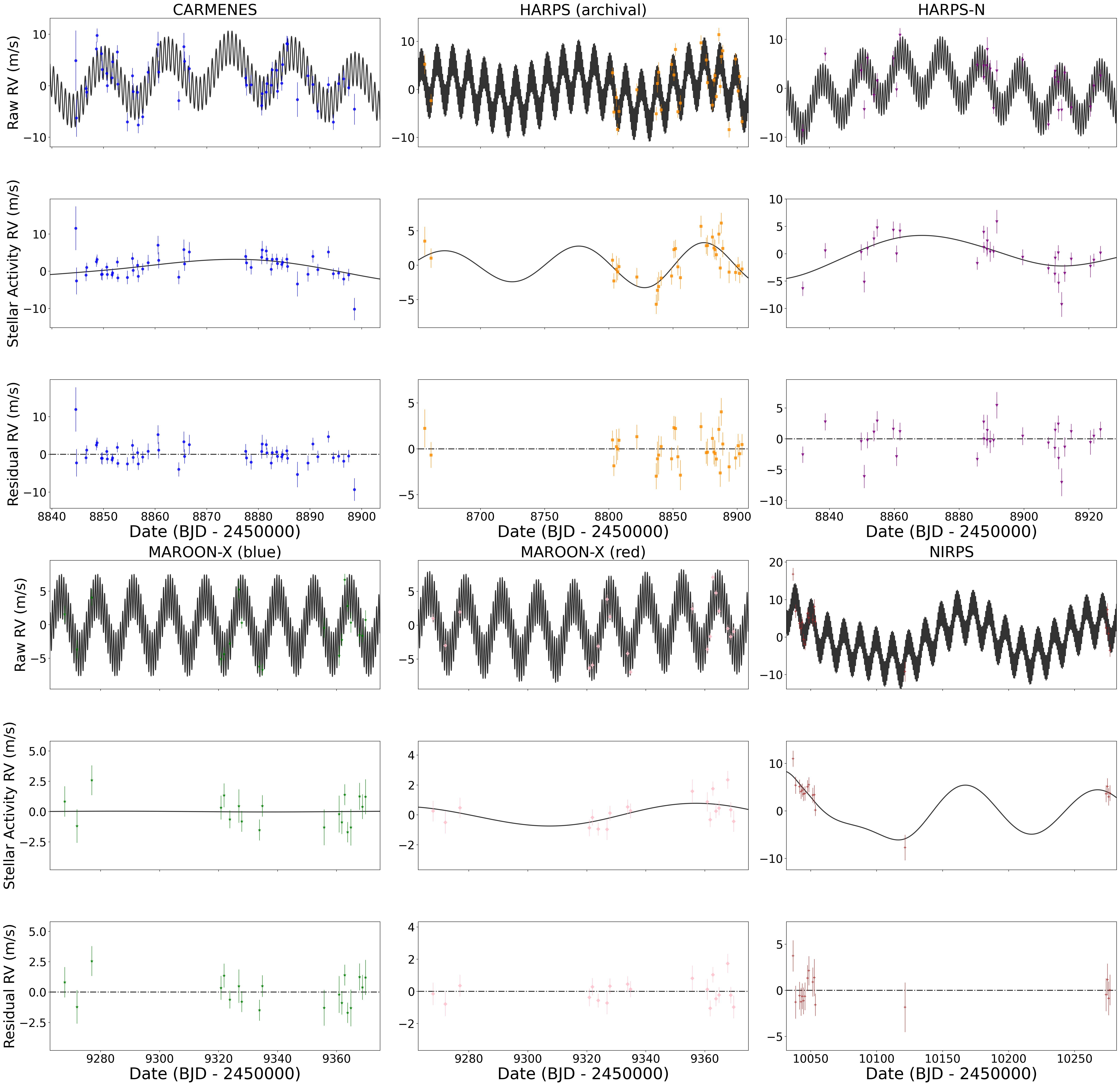}
    \caption{The RV time series used in the LTT\,3780 RV modeling. Each set of three rows shows the RV time series for a given instrument. The top row shows the mean-subtracted RV-time series for each instrument, with the maximum a-posteriori RV model shown in black, while the middle row shows just the stellar activity component of the RV time series, with the corresponding best-fit stellar activity signal shown in black. The residuals between the best-fit model and each time series are shown in the bottom row.}
    \label{fig:ltt3780-component}
\end{figure*}

\clearpage
\section{Stellar abundances} \label{appendix:abundances}

\begin{table*}[!h]
\caption{Reported stellar abundances}
\centering
    \begin{threeparttable}
    \begin{tabularx}{\linewidth}{C C C C C C C C C}
    \hline\hline
Abundance & \# Lines & [X/H] & $\sigma_\textrm{rand}$ & $\sigma_\textrm{Teff}$ & $\sigma_\textrm{[M/H]}$ & $\sigma_\textrm{vmic}$ & $\sigma_\textrm{vmac}$ & $\sigma_\textrm{total}$ \\
\hline
    & & & & \textbf{GJ 1132} \\
{[Fe/H]} & 22 & \textbf{-0.177} & $0.053$ & $0.032$ & $0.129$ & $0.031$ & $0.018$ & \textbf{0.148} \\
{[Mg/H]}\tnote{1} & 2 & \textbf{-0.295} & $0.054$ & $0.061$ & $0.070$ & $0.021$ & $0.024$ & \textbf{0.112} \\
{[Si/H]}\tnote{1} & 4 & \textbf{0.099} & $0.107$ & $0.096$ & $0.127$ & $0.001$ & $0.002$ & \textbf{0.192} \\
{[Ti/H]} & 53 & \textbf{-0.187} & $0.029$ & $0.041$ & $0.155$ & $0.028$ & $0.014$ & \textbf{0.166} \\
    \hline
    & & & & \textbf{GJ 1252} \\
{[Fe/H]} & 27 & \textbf{-0.176} & $0.037$ & $0.061$ & $0.058$ & $0.019$ & $0.087$ & \textbf{0.128} \\
{[Mg/H]} & 11 & \textbf{-0.035} & $0.068$ & $0.224$ & $0.026$ & $0.007$ & $0.032$ & \textbf{0.238} \\
{[Si/H]} & 5 & \textbf{-0.006} & $0.108$ & $0.360$ & $0.072$ & $0.002$ & $0.050$ &  \textbf{0.386}\\
    \hline
    & & & & \textbf{LTT 3780} \\
{[Fe/H]} & 38 & \textbf{0.061} & $0.024$ & $0.033$ & $0.079$ & $0.017$ & $0.104$ & \textbf{0.138} \\
{[Mg/H]} & 8 & \textbf{0.074} & $0.062$ & $0.132$ & $0.055$ & $0.002$ & $0.018$ & \textbf{0.157} \\
{[Si/H]} & 6 & \textbf{0.202} & $0.108$ & $0.189$ & $0.065$ & $0.002$ & $0.025$ & \textbf{0.229} \\
    \hline
    & & & & \textbf{GJ 357} \\
{[Fe/H]} & 44 & \textbf{-0.117} & $0.039$ & $0.039$ & $0.075$ & $0.011$ & $0.078$ & \textbf{0.122} \\
{[Mg/H]} & 6 & \textbf{-0.266} & $0.063$ & $0.063$ & $0.069$ & $0.010$ & $0.023$ & \textbf{0.115} \\
{[Si/H]} & 6 & \textbf{-0.261} & $0.070$ & $0.122$ & $0.065$ & $0.002$ & $0.054$ & \textbf{0.164} \\
    \hline
    & & & & \textbf{HD\,260655} \\
{[Fe/H]} & 72 & \textbf{-0.477} & $0.034$ & $0.033$ & $0.031$ & $0.013$ & $0.022$ & \textbf{0.062} \\
{[Mg/H]} & 7 & \textbf{-0.319} & $0.046$ & $0.060$ & $0.046$ & $0.014$ & $0.030$ & \textbf{0.094} \\
{[Si/H]} & 14 & \textbf{-0.568} & $0.023$ & $0.098$ & $0.047$ & $0.002$ & $0.023$ & \textbf{0.113} \\
    \hline
    & & & & \textbf{L 98-59} \\
{[Fe/H]} & 46 & \textbf{-0.287} & $0.045$ & $0.063$ & $0.109$ & $0.016$ & $0.062$ & \textbf{0.148} \\
{[Mg/H]} & 9 & \textbf{-0.336} & $0.058$ & $0.188$ & $0.051$ & $0.008$ & $0.026$ & \textbf{0.205} \\
{[Si/H]} & 6 & \textbf{-0.274} & $0.100$ & $0.208$ & $0.155$ & $0.006$ & $0.024$ & \textbf{0.279} \\
    \hline
    & & & & \textbf{LHS\,1140} \\
{[Fe/H]} & 17 & \textbf{-0.250} & $0.069$ & $0.027$ & $0.080$ & $0.012$ & $0.013$ & \textbf{0.110} \\
{[Mg/H]}\tnote{1} & 4 & \textbf{0.066} & $0.103$ & $0.091$ & $0.013$ & $0.001$ & $0.007$ & \textbf{0.138} \\
{[Si/H]}\tnote{1} & 1 & \textbf{0.570} & -- & $0.045$ & $0.014$ & $0.000$ & $0.006$ & \textbf{0.048} \\
{[Ti/H]} & 35 & \textbf{-0.304} & $0.042$ & $0.045$ & $0.101$ & $0.010$ & $0.016$ & \textbf{0.120} \\
    \hline
    & & & & \textbf{TOI-270} \\
{[Fe/H]} & 61 & \textbf{-0.142} & $0.032$ & $0.059$ & $0.060$ & $0.010$ & $0.072$ & \textbf{0.116} \\
{[Mg/H]} & 8 & \textbf{-0.312} & $0.047$ & $0.104$ & $0.060$ & $0.011$ & $0.026$ & \textbf{0.132} \\
{[Si/H]} & 6 & \textbf{-0.195} & $0.063$ & $0.164$ & $0.090$ & $0.001$ & $0.034$ & \textbf{0.191} \\
    \hline
    \end{tabularx}
    
    \smallskip
    \footnotesize

    This table details the results of the stellar abundance calculations discussed in this paper. $\sigma_\textrm{rand}$ refers to the standard deviation of the mean abundance of all the lines of a corresponding element. {The final elemental abundances and uncertainties are bolded, to distinguish them from the individual sources of error.}

    \begin{tablenotes}
    \item[1] The [Mg/H] and [Si/H] abundances for GJ 1132 and LHS\,1140 are not considered reliable, due to the low number of spectral lines used and their poor quality, and is reported here for completeness only; see Section \ref{subsec:stellchar-abund} for more details.
    \end{tablenotes}
    
    \end{threeparttable} \label{table:abundances-appendix}
\end{table*}

\end{document}